\documentstyle[11pt,aaspp4]{article}
\def\xxi0{\hbox{${\xi^\prime}_0$}}
\def\th0{\hbox{$\theta_0$}}
\def\r0{\hbox{$\varpi_0$}}

\def\f0{\hbox{$\psi_0$}}

\def\v1{\hbox{$v_{K1}$}}
\def\ro1{\hbox{$\rho_1$}}
\def\n1{\hbox{$n_1$}}
\def\b1{\hbox{$B_1$}}

\def\mstar{\hbox{$M_\ast$}}

\def\vecB{\hbox{$\vec{B}$}}
\def\bt{{\bf B}_t}
\def\msolyr{\hbox{${\rm M}_\odot\, {\rm yr}^{\rm -1}$}}
\def\msol{\hbox{${\rm M}_\odot $}}
\def\rsol{\hbox{${\rm R}_\odot $}}

\def\kelvin{{\rm K}}
\def\gauss{{\rm G}}

\def\kb{\hbox{$k_{\rm B}$}}

\def\pc{\hbox{${\rm pc}$}}
\def\ccc{\hbox{${\rm cm^{-3}}$}}

\def\function#1#2{\hbox{$#1\left(#2\right)$}}

\def\refeqm#1{(\ref{eq:#1})}
\def\refeqp#1{[\ref{eq:#1}]}

\def\subsectn#1{
    \addtocounter{subsection}{1}
    \setcounter{subsubsection}{0}
    {\ifnum\value{subsection}>1 \vskip 0.25 truein \penalty 10000 \fi}
    \penalty 10000
    {\centerline {\it \thesubsection.\ #1}}
    \penalty 10000      
    \bigskip
    \penalty 10000
    \index{#1}
    }
\def\subsubsectn#1{
    \addtocounter{subsubsection}{1}
    {\ifnum\value{subsubsection}>1 \vskip 0.125 truein \penalty 10000 \fi}
    \penalty 10000
    {\centerline {\it \thesubsubsection.\ #1}}
    \penalty 10000
    \bigskip
    \penalty 10000
    \index{#1}
    }

\renewcommand{\vec}[1]{{\mbox{{\boldmath$ #1 $}}}}
\def\bnabla{{\mbox{{\boldmath$ \nabla $}}}}
\def\bvecx{{\mbox{{\boldmath$\times$}}}}
\def\bvecdot{{\mbox{{\boldmath$ \cdot $}}}}

\newcommand{\al}{\mbox{ \raisebox{-.4ex}{$\stackrel{\textstyle <}{\sim}$} }}

\def\myr{\mbox{\rm Myr}}
\def\kni{\mbox{$k_{ni}$}}
\def\Mphi{\mbox{$M_\Phi$}}
\def\Mphio{\mbox{$M_{\Phi 0}$}}

\def\mcloud{\mbox{$M_0$}}
\def\mcore{\mbox{$M_{\rm c}$}}
\def\mc{\mbox{$m_{\rm c}$}}
\def\tad{\mbox{$t_{\rm AD}$}}
\def\tff{\mbox{$t_{\rm ff}$}}
\def\taucr{\tau_{\rm cr}}
\def\nuff{\nu_{\rm ff}}

\def\rhoc{\mbox{$\rho_{{\rm c},0}$}}
\def\bzero{\mbox{$B_0$}}
\def\rzero{\mbox{$R_0$}}
\def\rhat{\mbox{$\hat{r}$}}
\def\vhat{\mbox{$\hat{v}$}}
\def\rhohat{\mbox{$\hat{\rho}$}}
\def\dtfixm#1{\frac{D\, #1}{D t\,\,\,\,\,\,}}
\def\drfixt#1{\frac{\partial #1}{\partial r}}
\def\dtfixr#1{\frac{\partial #1}{\partial t}}
\def\dtaufixm#1{\frac{D\, #1}{D \tau\,\,\,\,\,}}
\def\dtaufixx#1{\frac{\partial #1}{\partial \tau}}
 
\tighten
\begin{document}
%
\title{Star Formation in Cold, Spherical, Magnetized Molecular Clouds}
\author{Pedro N. Safier\altaffilmark{1}, Christopher F.
McKee\altaffilmark{2},
and Steven W. Stahler\altaffilmark{3}}
\affil{University of California at Berkeley \\ Department of Astronomy\\
    Berkeley CA 94720-3411}
\altaffiltext{1}{Current address: University of Maryland at College
  Park, Department of Astronomy, College Park MD 20742; e-mail: {\it
    safier@astro.umd.edu\/}} 
\altaffiltext{2}{Physics Department, University of California at
Berkeley; {\it cmckee@astro.berkeley.edu}}
\altaffiltext{3}{\it stahler@astron.berkeley.edu}
%
\begin{abstract}
%
%
We present an idealized, spherical
model of the evolution of
a magnetized molecular cloud due to ambipolar diffusion.
This model allows us to
follow the quasi-static evolution of the cloud's core prior to collapse
and the subsequent
evolution of the remaining envelope. 
By neglecting the thermal pressure gradients in comparison with
magnetic stresses and by assuming that the ion velocity is small
compared with the neutral velocity,
we are able to find exact analytic solutions to
the MHD equations. We show that, 
in the case of a
centrally condensed cloud, a
core of finite mass collapses into the origin leaving behind a
quasi-static envelope, whereas initially homogeneous clouds never develop any
structure in the absence of thermal stresses, and collapse as a
whole. 
Prior to the collapse of the core,
the cloud's evolution is characterized by two phases: a long, quasi-static
phase where the relevant timescale is the ambipolar diffusion time
(treated in this paper),
and a short, dynamical phase where the characteristic timescale is the
free-fall time. The collapse of the core is an ``outside--in" collapse.
The quasi-static evolution terminates when the cloud 
becomes magnetically supercritical; thereafter its evolution 
is dynamical, and a  singularity develops at the origin---a protostar. 
After the initial formation of the protostar, the outer envelope
continues to evolve quasi-statically, while the region
of dynamical infall grows with time---an ``inside--out" collapse.
We use our solution to
estimate the magnetic flux trapped in the collapsing core and
the mass accretion rate onto the newly formed protostar.

Our results agree, within
factors of order unity, with the numerical results of Fiedler \& Mouschovias
\markcite{fm92}(1992)
for the physical quantities in the midplane of a collapsing, magnetized,
axisymmetric cloud up to the onset of dynamical collapse.  
Our simple approach thus captures the basic physics of a
self-gravitating,
magnetized cloud in which the evolution is driven by ambipolar diffusion.
It also enables us to treat the evolution of the accretion onto
the protostar after collapse, for which detailed numerical results
are as yet unavailable.
Remarkably, we 
find that at late times the accretion rate becomes comparable to that
of a non-magnetized, singular isothermal sphere, provided that
the ionization is due to galactic cosmic rays.

\end{abstract}


\newcounter{preprint}
\setcounter{preprint}{0}
\section{Introduction}
%
%
How do stars form? This  age-old question is one of the 
outstanding challenges of modern astrophysics.
Even though we still lack a definite answer to this question,
some partial answers have begun to emerge over the last three decades.
The birth of molecular radio astronomy and infrared astronomy
enabled us to definitively identify molecular clouds as the site of
star formation, and to probe the physical conditions therein
(Zuckerman \& Palmer, 1974\markcite{zp74}). Notwithstanding
the spectacular advances in instrumentation of the last decade and
the resulting explosive growth in the number of
 infrared and radio observations, the detection of a cloud undergoing
dynamical collapse is very difficult;
 to date there are only a few, tentative such detections 
(see Zhou et al., 1994\markcite{zho94} and references therein).
By necessity, then, one has to resort to theory to
understand  how the molecular gas ultimately condenses into a star.

The efforts in this direction can be traced back to
 the pioneering work of
Larson (1969\markcite{lar69}) and Penston (1969\markcite{pen69}).
 In these early calculations only thermal
pressure---and sometimes rotation--- were included, while magnetic
fields were neglected (see Mouschovias [1978]\markcite{mou78}
 and Bodenheimer \& Black [1978]\markcite{bod78}
 for a list of early references).  Of particular importance are the
semi-analytic similarity solutions of Shu (1977)\markcite{shu77}
 and Hunter (1977)\markcite{hun77}.
These spherical solutions, though idealized,
 shed important light on the nature of the non-magnetic collapse.
Non-magnetic collapse calculations are still pursued today;
the current state of development in this field is well represented by
the work of Myhill \& Boss (1993)\markcite{mb93},
Boss \& Myhill (1993)\markcite{bm93}, and Foster \& Chevalier
 (1993\markcite{fc93}).

The relevance of magnetic fields to star
formation were first pointed out by Mestel \& Spitzer
(1956)\markcite{ms56}. Using the virial theorem, 
Strittmater (1966)\markcite{\str66} calculated the
critical mass for gravitational collapse perpendicular to the field
lines of a cold, uniformly magnetized cloud, and Field (1965)\markcite{fld65}
 studied the effect of the magnetic field on the thermal instability. These
studies established that the presence of a magnetic field introduces
qualitative changes to the conclusions of non-magnetic studies.
However, not until the late seventies were detailed numerical calculations
of magnetic clouds  carried out. 
Mouschovias (1976a,b)\markcite{mou76a}\markcite{\mou76b} performed
the first self-consistent numerical studies of equilibrium,
self-gravitating magnetic clouds, and
Scott \& Black (1978)\markcite{sb80} were the first to study the dynamics of
collapsing, non-rotating clouds with a frozen-in magnetic field.
With the introduction of a
magnetic field, a realistic description  requires at least
two spatial dimensions. Moreover, at the densities of interest ($n\ga
10^3 {\rm cm^{-3}}$) the degree of ionization is low, and one has to
consider the ion and neutral material as two separate fluids.
This two-fluid nature of the problem has important consequences for the
stability and collapse of magnetized clouds. The drift of the neutrals
relative to the ions---a process known as ambipolar diffusion (Mestel
\& Spitzer, 1956)\markcite{ms56}---is responsible for
bringing to the verge of collapse an otherwise stable
cloud. As ambipolar diffusion proceeds, the density of the central
region grows until the inner core attains a mass larger than a
critical mass. Once this happens, the central region of the cloud is
in a state of dynamical collapse (for a review, see Mouschovias
[1996]\markcite{mou96} and McKee et. al [1993]\markcite{mz93}
and references therein).

Except for some idealized, one-dimensional solutions
which illustrated the role played by ambipolar diffusion in the
redistribution of magnetic flux  
(Mouschovias 1979\markcite{mou79}; Shu 1983\markcite{shu83};
 Scott, 1984\markcite{sco84}),
 the study of the stability and
collapse of realistic, magnetized clouds has been the domain of
complex numerical calculations, with major contributions by 
Nakano (1979\markcite{nak79}, 1982\markcite{nak82}, 1983\markcite{nak83});
Mouschovias and collaborators (Mouschovias, Paleologou, \& Fiedler
 1985\markcite{mp85}; Mouschovias
\& Morton, 1991\markcite{mm91}; Fiedler \& Mouschovias,
1992\markcite{fm92}; Basu \& Mouschovias, 1994\markcite{bm94},
1995a\markcite{bm95a},b\markcite{bm95b}; Ciolek \& Mouschovias,
 1993\markcite{mcm93}, 
1994a\markcite{cm94a},b\markcite{cm94b}, 1995\markcite{cm95});
Tomisaka, Ikeuchi, \& Nakamura
(1988a,b\markcite{tin88a}\markcite{tin88b};
 1989\markcite{tin89}; 1990\markcite{tin90}); Tomisaka, 
(1991{\markcite{tom91});
 Lizano \& Shu (1989)\markcite{ls89}; and
Galli \& Shu (1993a,b\markcite{gs93a}\markcite{gs93b}).
In contrast to the
non-magnetic models, none of these calculations 
can follow the cloud's evolution 
up to and beyond the point where a star is formed. They
either stop at an early stage of dynamical collapse (Fiedler \&
Mouschovias, 1992\markcite{fm92}); or tackle the 
problem {\it after} the formation of
the star and follow the infall of the envelope under the
gravitational field of a point source (Galli \& Shu,
1993b\markcite{gs93b}).

The shortcomings of current numerical work, and the lack of analytic or
semi-analytic models to complement and illuminate the numerical
results, motivated us to undertake the present study. Our goal is to
understand---by means of a very simple model---how a globally stable
cloud evolves, through ambipolar diffusion, into a state where a
central core collapses, and to shed some light
on the evolution of the remnant envelope. By doing so, we hope to
distill the basic features that can be lost in the detail and
complexity of numerical calculations.
Previous attempts along these lines adopted either planar 
(Shu, 1983\markcite{shu83}; Paleologou \& Mouschovias, 1983\markcite{pm83})
or cylindrical symmetry (Mouschovias \& Morton, 1991\markcite{mm91}). 
These clouds are always
magnetically subcritical under conditions of flux freezing, and
are of limited relevance to the study of collapsing clouds
(Mouschovias, 1991\markcite{mou91}; McKee et al., 1993\markcite{mz93}). 
In our case, we adopt {\it spherical} symmetry,
where the only non-vanishing component of the Lorentz force is the
scalar term $\propto dB^2/dr$. In contrast to the planar and
cylindrical cases, we cannot identify the actual magnetic field lines,
but a three-dimensional
cloud {\it can} undergo collapse even when flux freezing holds 
(Mouschovias, 1991\markcite{mou91}; McKee et al., 1993\markcite{mz93}). 

In this paper, we
neglect thermal pressure
altogether, and set the gas temperature equal to zero. The case
where thermal support is included is deferred to a second paper.
Our motivation
for this drastic approximation is twofold. From a practical viewpoint,
this simplification enables us to  
treat ambipolar diffusion from the onset of cloud contraction
to the collapse of a stellar core and beyond,
an intrinsically non self-similar problem.
Second, we show explicitly (\S 5.2.2) that the role of thermal pressure
diminishes in time as the central density rises.  Thus, our
calculation should give an increasingly accurate representation of the
cloud's later evolution, as it forms a protostar and then slowly drains
onto the protostar.

The paper is organized as follows: in \S 2 we introduce our basic
equations and approximations and present a general solution; the
general properties of the solution for a particular family of initial
conditions are presented in \S 3; in \S 4
we study the collapse of a homogeneous cloud; \S 5 contains our
results for a centrally condensed cloud and a comparison between our
results and the numerical results of Fiedler \& Mouschovias
(1992)\markcite{fm92}; and  our conclusions are presented in \S 6.

\section{Formulation and a General Solution}

We start with the general, two-fluid MHD equations governing ambipolar
diffusion in self-gravitating molecular clouds (see, e.g., Paleologou
\& Mouschovias, 1983\markcite{pm83}). We assume steady-state 
ionization---a situation
that obtains in dense clouds---and
adopt the ionization law
\begin{equation}
n_i = K_i\, n_n^{1/2}, \label{eq:ioneq}
\end{equation}
where $n_i$ and $n_n$ are, respectively, the ion and neutral number density
sand  $K_i$ is a constant with value
$K_i = 9.5\, \times 10^{-6}\, {\rm cm^{-3/2}}$ when 
HCO$^+$ is the most abundant ion (Elmegreen, 1979\markcite{elm79}).

Our first basic approximation is that  thermal
stresses are negligible compared to  magnetic forces, and we set the
temperature of the gas equal to zero. The physical regime where this
approximation is {\it likely}
 to be appropriate can be evaluated with the
help of  the parameter
\begin{equation}
\beta=\frac{8\pi \kb\,n_n\,T}{B^2}, \label{eq:eqbeta}
\end{equation}
where \kb\ is Boltzmann's constant, $T$ is the gas temperature, and $B$ is
the strength of the magnetic field. 
For typical conditions in a
starless cloud (see, e.g., Cernicharo, 1991\markcite{cer91}; 
Heiles, 1987\markcite{hei87}; Myers, 1985\markcite{mye85})
\begin{equation}
\beta = 0.04
\left(\frac{n_n}{10^3\,\ccc}\right)\left(\frac{T}{10\,\kelvin}\right) \left(\frac{B}{30\,\mu\gauss}\right)^{-2}.\label{eq:quantbeta}
\end{equation}
Therefore, low-mass clumps are magnetically dominated prior to the
point at which their density is high enough for them to be visible as
ammonia cores. The line widths of more massive clumps are often
supersonic, so thermal pressure is relatively small there as
well. However, we have not explicitly included turbulent pressure in
our models because, in this formulation,
 its effects could be incorporated only under the
artificial assumption that it is proportional to the static magnetic pressure.

To further simplify the problem
we impose spherical symmetry. 
The accuracy of this approximation is unclear.
Theoretical models of cloud cores suggest
that the cores are flattened along field lines (see, eg., Fiedler \&
Mouschovias, 1992\markcite{fm92}; Lizano \& Shu, 1989\markcite{ls89};
Tomisaka, Ikeuchi, \& Nakamura, 1988a\markcite{tin88a}).
The observational evidence on cloud shapes has been interpreted as
indicating that they are prolate (Myers et. al, 1991\markcite{mfgb}; 
Ryden, 1996\markcite{ryd96})  or
that they are indeed flattened along field lines but that they are
toroidal (Li \& Shu 1996\markcite{ls96}).  Physically,
the thermal and turbulent pressure, which we have neglected, prevent
actual cloud cores from becoming too flattened, and therefore we
expect that our model will be qualitatively correct. Spherically
symmetric models of magnetized clouds have been shown to represent the
conditions for the onset of gravitational collapse reasonably well
(Mouschovias \& Spitzer, 1976\markcite{ms76}; Holliman \& McKee,
1996\markcite{hm96}). Because we impose spherical symmetry, 
we cannot identify the actual
magnetic field lines, nor can we account for nonradial forces. 
However, since we shall follow only
the quasi-static part of the collapse (see below), 
we need the force equation only
to estimate the flux remaining in the cloud, not to determine
the dynamics of the collapse. 
As a result, we can adopt a simple approximation for
the magnetic force, keeping the pressure gradient associated
with the tangential fields but ignoring the associated
tension.  Let $\bt\equiv \vecB_\theta+\vecB_\phi$ 
be the tangential field.  The radial component of the magnetic
force is then
\begin{equation}
F_r=\frac{1}{4\pi}\left(-\frac12\frac{\partial B_t^2}{\partial r}
        +\bt\bvecdot\bnabla B_r-\frac {B_t^2}{r}\right).
\label{eq:force}
\end{equation}
The last term represents the hoop stress, an inward force
due to magnetic tension.  However, in an open field configuration,
which we assume, this force is more than compensated by the second
term, which represents an outward tension.  
Since this second term cannot be estimated within the 
context of a spherical model, we simply ignore the last two terms,
representing tension. Numerical simulations (cf. Fiedler \&
Mouschovias, 1992; Lizano \& Shu, 1989) show that during the
quasistatic phase the field does not develop a strong
curvature. Therefore, our neglect of tension should be accurate within
a factor $\sim 2$.

        To repeat, then, we ignore the last two terms in 
equation~(\ref{eq:force})
 and keep the first term, representing pressure. We eliminate the
 angular dependence of $B_t^2$ by using its angular average
(cf. Chiueh \& Chou 1994\markcite{cc94});
and, for  simplicity, we shall use the notation $B$ instead of $\langle
B_t^2\rangle^{1/2}$, where $\langle\,\rangle$ stand for the angular average.

        Finally, at the typical densities of dense molecular clouds the
ionization fraction is very small, and therefore, to a good degree of
approximation, we have that
\begin{equation}
\rho_i \left(\frac{\partial v_i}{\partial t} +
             v_i\,\frac{\partial v_i}{\partial r}\right)
 \approx 0\label{eq:11a}
\end{equation}
and
\begin{equation}
  \label{eq:11b}
 \frac{ G M \rho_i}{r^2} \approx 0.
\end{equation}

        With these assumptions of negligible thermal pressure and of spherical
symmetry, and neglecting the inertia of the ions, the
momentum equations for the neutral and ion fluids and the mass
conservation equation for the neutrals take the form
\begin{eqnarray}
\rho_n \dtfixm{v_n} & = & \rho_i\, \rho_n\,\frac{\left<\sigma
                          v\right>_{ni}}{m_i} v_D - 
                              \frac{ G M
                              \rho_n}{r^2}, \label{eq:01} \\
                  0 & = & 
                          -\rho_i\, \rho_n\,\frac{\left<\sigma
                          v\right>_{ni}}{m_i} v_D
                              -\frac{1}{8\pi} \drfixt{B^2}   \label{eq:02}
\end{eqnarray}
and
\begin{equation}
\dtfixm{\rho_n}  =  - \frac{\rho_n}{r^2} 
                \drfixt{\left(r^2\,v_n\right)}, \label{eq:3}
\end{equation}
where 
\begin{equation}
  \label{eq:ddt}
  \frac{D\,\,}{Dt} \equiv \frac{\partial }{\partial t} +
                          v_n\,\frac{\partial}{\partial r}
\end{equation}
indicates a substantial derivative;
$v_D\equiv v_i-v_n$ is the ion-neutral drift velocity; 
$\left<\sigma v\right>_{ni}$
is the average collision rate between neutrals and ions, with
the value $1.7\times 10^{-9}\,\ccc\,{\rm s^{-1}}$
for HCO$^+$--H$_2$ collisons (McDaniel \& Mason, 1973\markcite{mm73});
and all the other
symbols have their usual meanings, with the subscripts $i$ and $n$
standing, respectively, for ion and neutral quantities.
In writing equations (\ref{eq:01}) and (\ref{eq:02}), we have
neglected the neutral mass in comparison with the ion mass,
$m_n\ll m_i$.
We define a new constant $k_{ni}$ by
\begin{equation}
\kni \equiv  K_i \frac{\left<\sigma v\right>_{ni}}{{m_n}^{1/2}}
        =8.2\times 10^{-3}\, {\rm cm^{3/2}\, s^{-1}\, g^{-1/2}}\,, 
\label{eq:kinv}
\end{equation}
where $m_n$ and $m_{\rm H}$ are, respectively, the neutral's and
proton's mass,  and where we have set $m_n = 2.33\, m_{\rm H}$.

        At this point it is important to make contact with previous work by
other authors. We choose to focus on the solutions  by
Mouschovias and collaborators 
(see references in \S 1) because our approach is closest to theirs.
 Their solutions depend on three
dimensionless parameters, one of which is (Fiedler \& Mouschovias, 1992)
\begin{equation}
  \label{eq:nuff}
  \nuff  = \left(\frac{8}{3\pi^2}\right)^{1/2}\,\,\frac{t_{\rm ff,0}}{t_{ni,0}},
\end{equation}
where 
\begin{equation}
  \label{tff0}
  t_{\rm ff,0} = \left(\frac{3\pi}{32G\rho_{n,0}}\right)^{1/2}
\end{equation}
is the free-fall time for their initial reference state with uniform
neutral density $\rho_{n,0}$, and where
\begin{equation}
  \label{eq:tin0}
  t_{ni,0} = \frac{1}{n_{i,0}{\left<\sigma
               v\right>_{ni}}}
\end{equation}
is the neutral-ion collision time for the initial reference state
(note that $\nuff $ depends only on universal constants and the
microphysics of the problem, and that we have again assumed
$m_n\ll m_i$).
Our parameter $\kni$ is directly related to
$\nuff $ through
\begin{equation}
  \label{eq:nuff0kin}
\kni   = 3.54\,G^{1/2}\,\nuff .
\end{equation}

        Using  $\kni$ and equation \refeqm{ioneq}, 
the system of eqs. \refeqm{01} and \refeqm{02} take the form
\begin{eqnarray}
\rho_n \dtfixm{v_n} & = & \kni \rho_n^{3/2} v_D - 
                              \frac{ G M
                              \rho_n}{r^2}, \label{eq:1} \\
                  0 & = & -\kni \rho_n^{3/2} v_D - 
                              \frac{1}{8\pi} \drfixt{B^2}
                                                        \label{eq:2} 
\end{eqnarray}

In principle, one has to add to eqs. \refeqm{3}, \refeqm{1}, and
\refeqm{2} the
induction equation to solve self-consistently for the magnetic field.
We postpone our discussion of the induction equation until we consider other
approximations that apply to the solution of eqs.
\refeqm{3}, \refeqm{1}, and \refeqm{2};
it turns out that, to the same order of approximation
for which our solution of eqs. \refeqm{3}, \refeqm{1}, and \refeqm{2}
 holds, the induction equation is automatically satisfied to
 sufficient accuracy.

Our fundamental approximations to solve eqs. \refeqm{3}, \refeqm{1},
and \refeqm{2} are two. 
First, we assume that the evolution is quasi-static, and set
\begin{equation}
\dtfixm{v_n}\approx 0. \label{eq:8}
\end{equation}
Second, we assume that the ions are effectively static, so that
their velocity is small compared even with that of the neutrals:
\begin{equation}
v_i \ll v_n, \label{eq:9}
\end{equation}
so that
\begin{equation}
v_D \approx -v_n. \label{eq:10}
\end{equation}

        These approxmations require comment.  Recall that
we are interested in the case where only the magnetic field
supports the cloud against gravity. Because only the ions are directly
coupled to the field, the neutrals are subject to the magnetic forces
only through collisions with the ions, and there is a net slip of the
neutral material past the magnetic field lines
(Mestel \& Spitzer, 1956). 
If this collisional coupling is effective, 
the drift velocity $v_D$ will be
small.  There are two ways in which this can occur: 
On the one hand, when $v_i
\sim v_n$, the drift velocity is small, but the ions are essentially
comoving with the neutrals; in other words, the cloud is not
magnetically supported and it undergoes dynamical collapse
(Mestel \& Spitzer, 1956; 
note that when $v_i=v_n$ the
first term on the rhs of equation \refeqp{1} vanishes, and the material is
in free-fall). On the other hand, $v_D$ is small also when $v_i \ll
v_n$, and $v_n$ itself is small. In this case, the ions---and the
magnetic field lines---are essentially stationary, and the neutral
material slowly drifts by them. Thus,
provided that $v_n$ is much smaller than the local free-fall velocity,
the coupling between the ions and the neutrals is good, and the
magnetic field is effective in supporting the cloud on timescales of
order the free-fall time, {\it but on a longer timescale the cloud is
in a state of quasistatic contraction\/} 
(e.g., Lizano \& Shu 1989; 
Mouschovias, 1991). This
is the case we consider here:
quasi-static contraction of the neutrals through essentially static
ions.

        Using eqs.~\refeqm{8} and \refeqm{10}, equation~\refeqm{1} becomes
\begin{equation}
v_n = - \frac{GM}{\kni\, r^2 \rho_n^{1/2}}. \label{eq:12a}
\end{equation}
Substitution into equation~\refeqm{3} then gives
\begin{equation}
\dtfixm{\rho_n}=\frac{G}{\kni} \frac{\rho_n}{r^2}
\frac{\partial}{\partial r}\,
\left(\frac{M}{\rho_n^{1/2}}\right).\label{eq:12b}
\end{equation}

Lagrangian coordinates are better suited for the solution of equation
\refeqm{12b}. By using the transformation
\begin{equation}
\frac{\partial r}{\partial M}\,
 = \frac{1}{4\pi r^2\rho_n} \label{eq:13}
\end{equation}
on equation~\refeqm{12b}, we finally get our fundamental equation:
\begin{equation}
\dtfixm{\rho_n}=\frac{4\pi G}{\kni} \rho_n^2 
\frac{\partial}{\partial M}\,
\left(\frac{M}{\rho_n^{1/2}}\right).\label{eq:14}
\end{equation}

Note that equation \refeqm{14} is a self-contained equation for
$\rho_n(t,M)$; once a solution is found, $v_n$ and $r$ can be found
from eqs. \refeqm{12a} and \refeqm{13}, respectively, while $B$ is obtained by
quadrature of
\begin{equation}
\frac{\partial B^2}{\partial M}\,
 = - \frac{2 G M}{r^4}, \label{eq:15}
\end{equation}
which follows from eqs. \refeqm{1} and \refeqm{2} in the limit where
eqs. \refeqm{11a}, \refeqm{11b}, and \refeqm{8}  are valid.

\subsection{Self-Consistency and the Induction Equation} 
{}

One can check, {\it a posteriori}, the validity of the quasi-static
approximation (eq. \refeqp{8}) in the following way. 
For a given solution, evaluate the Lagrangian time derivative of
$v_n$. Then, in terms of the parameter $\alpha$ defined by
\begin{equation}
 \dtfixm{v_n}  \equiv  -\alpha\,\frac{GM}{r^2(M)}, 
\label{eq:alphadef}
\end{equation}
the quasi-static approximation is valid as long
as $\alpha \ll 1$.

Our second approximation in solving eqs. \refeqm{3},\refeqm{1}, and
 \refeqm{2} is
given by equation \refeqm{9}, i.e., that the motion of the ions is negligible
compared to that of the neutral material. The magnitude of $v_i$ is
critical to the solution of the induction equation,
\begin{equation}
\frac{\partial\vec{B}}{\partial t} = 
 \bnabla \bvecx \biggl( \vec{v_i}\bvecx\vec{B}\biggr).
                   \label{eq:induc1}
\end{equation}
In the limit where $v_i=0$ identically, the magnetic field is constant
in time, and the neutrals simply move across the field lines, which
are fixed in space. In this case, the spatial gradient of the field is
given by the condition of hydrostatic equilibrium, equation \refeqm{15},
and the induction equation is automatically satisfied. This simple
situation 
still applies for finite $v_i$
as long as $v_i \ll v_n$. To quantify matters, we rewrite the induction
equation heuristically as
\begin{equation}
  \label{eq:fluxfreeze1}
\dtfixr{\Phi}+ v_i \drfixt{\Phi} = 0,
\end{equation}
where $\Phi$ is the magnetic flux, defined by
\begin{equation}
  \label{eq:fluxdef}
  \drfixt{\Phi} = 2\pi B r.
\end{equation}
Equation \refeqm{fluxfreeze1} can be rewritten as
\begin{equation}
  \label{eq:fluxfreeze2}
  \dtfixr{\Phi}= -2\pi\,\frac{v_i}{v_n}\,v_n\,Br.
\end{equation}
Let $L$ be a characteristic length of the problem, and $t_n$ a
characteristic time for the motion of the neutrals; then
\begin{equation}
  \label{vnorder}
  v_n \sim \frac{L}{t_n},
\end{equation}
and to within order of magnitude equation \refeqm{fluxfreeze2} reads
\begin{equation}
  \label{eq:fluxfreezeorder}
  \dtfixr{\Phi} \sim  \frac{v_i}{v_n}\,\frac{BL^2}{t_n} 
                   \sim  \frac{v_i}{v_n}\,\frac{\Phi}{t_n}.
\end{equation}
Equation \refeqm{fluxfreezeorder} shows that
the characteristic time for the magnetic flux to change is
a factor $v_n/v_i$ longer than the characteristic time for the motion
of the neutrals. Therefore, as long as $v_i/v_n \ll 1$ the induction
equation need not be solved separately, 
and the strength of the magnetic field is
given by the condition of hydrostatic equilibrium. This means that the
magnetic and hydrodynamic parts of the problem decouple, resulting in
an enormous simplification.

The validity of the condition $v_i/v_n \ll 1$ can be checked {\it a
  posteriori\/} by  solving for $v_i$ from equation \refeqm{fluxfreeze1},
\begin{equation}
  \label{eq:vicheck}
  v_i = -\frac{\partial \Phi/\!\partial t }
      {\partial \Phi/\!\partial r },
\end{equation}
once $B$ and $\Phi$ are found, respectively, from eqs. \refeqm{15}
and \refeqm{fluxdef}.

Finally, the assumption that the thermal stresses are negligible
compared to the magnetic stresses can be checked {\it a posteriori\/} by
evaluating a modified form of the parameter $\beta$ in equation~\refeqm{eqbeta}:
\begin{equation}
  \beta^\prime=8\pi\kb\,T\,\frac{\!\partial n/\partial r}
                            {\!\partial B^2/\partial r}.
                            \label{eq:betaprime}
\end{equation}

\subsection{Dimensionless Expressions}

It is convenient to cast all the equations in dimensionless form.
Let
\begin{eqnarray}
\rho_n\left(t,M\right) & = & \rhoc\, \rhohat \left(\tau,m\right), 
        \label{eq:18}\\
r\left(t,M\right) & = & \rzero\, \rhat\left(\tau,m\right), \label{eq:19}
\end{eqnarray}
and
\begin{equation}
B\left(t,M\right)  =  \bzero\, b\left(\tau,m\right), \label{eq:20}
\end{equation}
where \rhoc, \rzero, and \bzero\ are, respectively, the initial
central density, the initial radius of the cloud, and the magnetic
field at the edge of the cloud, and where we have
introduced a dimensionless time $\tau$ and a dimensionless mass $m$:
\begin{eqnarray}
t & = & \tad\, \tau, \label{eq:21} \\
M & = & \mcloud\, m. \label{eq:22}
\end{eqnarray}
Here \mcloud\ is the total mass of the cloud, and \tad\ is a
characteristic ambipolar diffusion timescale,
\begin{equation}
\tad  \equiv \frac{\kni}{4\pi G \rho_{c,0}^{1/2}}
     = 4.96\times 10^6\left(\frac{n_{c,0}}{10^3\,\ccc}
        \right)^{-1/2}~~{\rm yr},
\label{eq:23}
\end{equation}
where $n_{c,0}$ is the initial number density at the center of the cloud.
Note that the ratio $\tad/\tff$, where $\tff$ is the
characteristic free-fall time for the inital state, is given by
\begin{equation}
  \label{eq:tadtff}
  \frac{\tad}{\tff}  =  0.147\,\frac{\kni}{G^{1/2}} 
        = 4.7,
\end{equation}
or, in terms of the quantity $\nuff $ 
used by Mouschovias and collaborators,
\begin{equation}
  \label{eq:tadtffnu}
\frac{\tad}{\tff}=\left(\frac{8}{3\pi^2}\right)^{1/2}\,\nuff .  
\end{equation}

        Let us define the parameter $\epsilon$, which is inversely 
related to the initial degree of concentration of the cloud:
\begin{equation}
  \label{eq:epsilondef}
  \epsilon    \equiv  \frac{\mcloud}{\frac{4\pi}{3}\rzero^3\,\rhoc}
              =  \frac{\left<\rho\right>_0}{\rhoc}.
\end{equation}
Also, let $\vhat_i$ and $\vhat_n$ be, respectively, the
dimensionless ion and neutral velocities, defined to be positive,
\begin{equation}
v_n  \equiv  -v_{n,0}\, \vhat_n \label{eq:24a}
\end{equation}
and
\begin{equation}
v_i  \equiv  -v_{n,0}\, \vhat_i. \label{eq:24b}
\end{equation}
Here,
\begin{equation}
v_{n,0}  \equiv  \frac{G \mcloud}{\kni \rzero^2 \rhoc^{1/2}}
         =  \epsilon\frac{\rzero}{3 \tad} \label{eq:25}
\end{equation}
is a characteristic velocity of the neutrals.
Note that the initial velocity of the neutrals at the edge of the cloud is
$v_n=v_{n,0}[\rho_{c,0}/\rho_0(R_0)]^{1/2}\geq 1$.

        To characterize the magnetic field, we 
define a parameter $\delta$ as the initial 
ratio of gravitational to
magnetic pressure at the edge of the cloud:
\begin{equation}
\delta    \equiv \frac{2GM^2_{0}}{\rzero^4\,\bzero^2}.\label{eq:33}
\end{equation}
The magnetic field can also be characterized in terms of the magnetic
critical mass \Mphi, defined by
\begin{equation}
  \label{eq:mphidef}
  \Mphi = \frac{c_\Phi}{G^{1/2}}\,\Phi,
\end{equation}
where $c_\Phi$ is a numerical constant that
depends on the distribution of the mass to flux ratio
in the cloud; for a uniform spherical cloud with a constant
magnetic field, $c_\Phi\simeq 0.12$ (Tomisaka et al. 1988b). 
Clouds with $M > \Mphi$ (magnetically supercritical)
are dynamically unstable and will 
collapse even if the magnetic
field is perfectly coupled to the neutral material. 
In terms of the characteristic value for the initial magnetic critical
mass $M_{\Phi 0}  \equiv {c_\Phi} \pi R_0^2\bzero/G^{1/2}$, we have
\begin{equation}
  \label{eq:critratio0}
   \frac{\Mphio}{M_0} = {  \pi c_\Phi\surd 2\over \delta^{1/2}}
        =\frac{0.53}{\delta^{1/2}}.
\end{equation}
We shall focus on clouds that have small 
values of $\delta$, corresponding to 
magnetically subcritical clouds ($\Mphio/M_0>1$)
in which magnetic forces are initially stronger
than gravitational forces.

        Using the definitions in eqs.\refeqm{18}-\refeqm{33},
our basic equations \refeqm{13}, \refeqm{14}, and \refeqm{15} take the form
\begin{eqnarray}
\dtaufixm{\rhohat} & = & \rhohat^2 \frac{\partial}{\partial
                    m}\, 
                    \left(\frac{m}{\rhohat^{1/2}}\right), \label{eq:26} \\
\frac{\partial \rhat^3}{\partial m}\, & = & 
\frac{\epsilon}{\rhohat},
\label{eq:27} \\ 
\frac{\partial b^2}{\partial m} & = & -
                \delta\,\frac{m}{\rhat^4},\label{eq:28}
\end{eqnarray}
and equation \refeqm{12a} now reads
\begin{equation}
  \label{eq:29}
  \vhat_n  =  \frac{m}{\rhat^2\,\rhohat^{1/2}}. \label{29}
\end{equation}

The solution of equation \refeqm{26} requires the specification of an
initial condition $\rhohat_0\equiv\rhohat\left(\tau=0,m\right)$.
In addition, equations
\refeqm{27} and \refeqm{28}
 are subject to the boundary conditions
\begin{eqnarray}
\rhat\left(\tau=0,m=1\right)  & = &  1,  \label{eq:rhatbc1} \\
\left.\dtaufixx{\rhat}\right|_{m=0}  & = &  0, \label{eq:rhatbc2}
\end{eqnarray}
and
\begin{equation}
b\left(\tau,m=1\right)  =  1, \label{eq:28b}
\end{equation}
respectively. The boundary conditions in
eqs.~\refeqm{rhatbc1}-\refeqm{rhatbc2} are  obvious,
while the boundary condition given by equation \refeqm{28b} means that the
magnetic field at the cloud's edge is constant. This boundary condition
is appropriate for a dense cloud surrounded by a low density medium
threaded by a field $B_0$.

As an {\it a posteriori} check, the assumption of quasi-static
motion can be quantified with the help of the quantity $\alpha$
defined in equation~\refeqm{alphadef}
\begin{eqnarray}
\alpha & = & \frac{4\pi G}{\kni^2} \frac{\rhat^2}{m}\dtaufixm{\vhat_n}
\nonumber \\
       & = & 1.2\times 10^{-2}\,
         \frac{\rhat^2}{m}\dtaufixm{\vhat_n}, \label{eq:alphacalc}
\end{eqnarray}
where we used the value of $\kni$ in equation \refeqm{kinv}; the ratio
$v_i/v_n$ is given by
\begin{equation}
  \label{eq:vivncomp}
\frac{v_i}{v_n} = \frac{3}{\epsilon }\, \frac{1}{\hat{v}_n} \,
\frac{ \partial \phi/\!\partial \tau }
      {\partial \phi/\!\partial \hat{r} },
\end{equation}
where $\phi$ is the dimensionless magnetic flux defined by
\begin{equation}
  \label{eq:phidef}
  \Phi = \pi B_0\, R_0^2\, \phi;
\end{equation}
and, finally, by defining a central value of the parameter $\beta$ in
equation~\refeqm{eqbeta}, we can quantify the role of thermal stresses with
\begin{eqnarray}
  \label{eq:betaprimecomp}
  \beta^\prime & = &\frac{8\pi\kb\,n_{{\rm c},0}\,T}{B_0^2}\,
\frac{\partial \rhohat/\!\partial m }
      {\partial b^2/\!\partial m } \nonumber \\
               & \equiv & \beta_0\,
      \frac{\partial \rhohat/\!\partial m }
      {\partial b^2/\!\partial m },
\end{eqnarray}
where $n_{c,0}$ is the initial number density at the center of
the cloud.

To summarize, by putting our problem in dimensionless form,
we have reduced the dependence of the solution to only
two free parameters, $\epsilon$ and
$\delta$, together with a specification of the initial 
mass distribution in the cloud.
The parameter $\epsilon$ is the inverse of the initial degree
of concentration of the cloud (eq.~\refeqp{epsilondef}), and the
evolution of the hydrodynamical variables \rhat, \vhat, and \rhohat\
depends only on this parameter. The second free parameter, $\delta$
(eq.~\refeqp{33}), 
is the initial ratio of gravitational to magnetic pressure at the edge
of the cloud, and parameterizes the initial mass-to-flux ratio of the
cloud (eq.~\refeqp{critratio0}). The evolution of the magnetic field
and the ion velocity depends on $\delta$ and $\epsilon$. Also, we will
show below (eq.~\refeqp{taucr}) that
the value of $\delta$ determines the regime where our static-ion approximation
is valid.  
To convert our dimensionless solution into physical units,
it is necessary to specify the cloud mass $M_0$, the initial cloud
radius $R_0$, the magnetic field at the edge of the cloud $B_0$, and
the ambipolar diffusion time $\tad$ (eq. [\ref{eq:23}]) or the initial
 density at the center of the cloud $\rho_{c,0}$.

\subsection{A General Solution}

The key to the solution of the system of equations
\refeqm{26}-\refeqm{29} is the solution of equation \refeqm{26}. This can
easily be achieved by defining the new dependent variable
\begin{equation}
\chi \equiv \frac{m}{\rhohat^{1/2}} \label{eq:36}
\end{equation}
and the new independent variable
\begin{equation}
\xi \equiv \frac{1}{m}. \label{eq:37}
\end{equation}
Using these transformations, equation \refeqm{26} takes the standard form
\begin{equation}
a\left(\chi\right) \frac{\partial \chi}{\partial \xi} + 
\frac{\partial \chi}{\partial \tau} = 0, \label{eq:38}
\end{equation}
where
\begin{equation}
a\left(\chi\right)=-\frac{1}{2\chi}. \label{eq:39}
\end{equation}

        Equation \refeqm{38} is a one-dimensional conservation law:
if $\chi_0(\xi) \equiv  \chi\left(\tau=0,\xi\right)$
is the initial condition for $\chi$,
then the solution of equation \refeqm{38} is given by
\begin{equation}
\chi\left(\tau,\xi\right)=\chi_0\left[\xi-a\left(\chi\right)\tau\right].
\label{eq:41}
\end{equation}

Two things are remarkable about the solution given by equation \refeqm{41}.
First, it represents an implicit equation for $\chi$, and one hopes
that for certain forms of the initial condition $\chi_0\left(\xi\right)$
this equation can be solved algebraically. Second, note that the
solution for $\chi$ 
has the form of a wave with velocity \function{a}{\chi}, and therefore
there exists the possibility of wave steepening and shock formation,
i.e., the formation of a singularity.

        With $a(\chi)$ given by equation (\ref{eq:39}),
the solution in equation~\refeqm{41} becomes
\begin{equation}
  \label{eq:41b}
  \chi\left(\tau,\xi\right) = \chi_0\left(\psi\xi\right),
\end{equation}
where
\begin{equation}
  \label{eq:41c}
  \psi\equiv 1+ \frac{\tau}{2}\rhohat^{1/2}.
\end{equation}
Because $\chi=\left(\rhohat^{1/2}\xi\right)^{-1}$, we have 
the solution for the density in terms of its initial distribution:
\begin{equation}
  \label{eq:41d}
  \rhohat\left(\tau,\xi\right)=\psi^2\,\rhohat_0\left(\psi\xi\right).
\end{equation}

        To proceed, we must select a specific form for the inital density
distribution $\rhohat_0(\xi)$. A form that roughly approximates models
for molecular cores is
\begin{equation}
\rhohat_0=\frac{m_c^p}{\left(\mc+m\right)^p}
=\left(1+\frac{1}{m_c\xi}\right)^{-p},\label{eq:42}
\end{equation}
with \mc\ a constant. We shall concentrate on two cases: $p=2$, which
approximates the density profile of a 
{\it non-singular} isothermal cloud, and
$p=0$, which provides useful insights into the evolution of the
central region.
First, however, we consider some features of the general case.

\section{The Case of Arbitrary $p$}

Because
\begin{equation}
  \label{eq:3.1}
  m= M_0^{-1}\int_0^r\, 4\pi r^2\,\rho\,dr,
\end{equation}
equation~\refeqm{42} is an integral equation for
$\rhohat_0\left(\rhat\right)$, the initial density as a function of
radius. This equation is readily solved to give
\begin{equation}
  \label{eq:3.2}
  \rhohat_0 = \biggl[1+\frac{\left(p+1\right)\rhat^3}{\epsilon\, m_c}\biggr]^{-\frac{p}{p+1}}.
\end{equation}
Inserting this result back into equation~\refeqm{42} gives
\begin{equation}
  \label{eq:3.3}
 \frac{m}{\mc} = \biggl[1+\frac{\left(p+1\right)\rhat^3}{\epsilon\, m_c}
       \biggr]^{\frac{1}{p+1}} -1~~~~~~(\tau=0),
\end{equation}
and because, initially, $m=1$ at $\rhat=1$, this result provides the
relation between $\epsilon$ and \mc\ for our model,
\begin{equation}
  \label{eq:3.4}
  \epsilon =
        \frac{p+1}{\mc\,\biggl[\left({\displaystyle 1+\frac{1}{m_c}}
        \right)^{p+1}\,-1\biggr]}\,.
\end{equation}

We now evaluate the solution given by equation~\refeqm{41d} for the initial
density distribution we have adopted,
\begin{equation}
  \label{eq:3.5}
  \rhohat = \psi^2\left(1+\frac{m}{m_c\,\psi}\right)^{-p}.
\end{equation}
Using equation~\refeqm{41c} for $\psi$, 
we obtain the general solution
\begin{equation}
  \label{eq:3.6}
\frac{m}{m_c} = \left(1+\frac{\tau}{2}\,\rhohat^{1/2}\right)\, 
 \Biggl[\left(\frac{1}{\rhohat^{1/2}} + \frac{\tau}{2}\right)^{2/p}\,-1\Biggr].
\end{equation}

        We obtain the central density $\rhohat_c$ by setting $m=0$,
\begin{equation}
  \label{eq:3.7}
  \rhohat_c = \frac{4}{\left(2-\tau\right)^2}.
\end{equation}
Thus, the solution becomes singular at $\tau=2$. Near the center of
the cloud, the density becomes very large as $\tau\rightarrow 2$.
We can expand equation~\refeqm{3.6} as this limit is approached and find
\begin{equation}
  \label{eq:3.8}
  \frac{m}{m_c}\rightarrow \frac{2}{p}\left[1+\frac{2+p}{2 p\,
  \rhohat^{1/2}}\right]\,. 
\end{equation}
We conclude that at $\tau=2$ a core of mass $(2/p)m_c$ collapses to a
singularity. This behavior is an artifact of our neglect of thermal
pressure. Had we included thermal pressure, the singularity would have
been avoided, and instead a high density object---a protostar---would 
have formed at $\tau\approx 2$.

        One other feature of the solution at $\tau=2$ is worth
noting. Equation \refeqm{3.8} shows that the mass just outside the
singularity scales as $\rhohat^{-1/2}$. Assuming a power law for for
$\rhohat(\rhat)$ there, we find $\rhohat(\tau=2,\rhat)\propto
\rhat^{-2}$: at the instant the singularity forms, the density {\it
  just outside\/} has the same radial dependence as a singular
isothermal sphere. However, this gas is not static.

The conditions for the solution found here to be valid are more easily
discussed for specific models, so we now consider the collapse of a
homogeneous cloud.

\section{The Homogeneous Cloud}

The case of an initially homogeneous cloud ($p=0$, $\epsilon=1$)
 is of particular
interest because a complete solution can be obtained analytically,
 and because it gives
some insights into how the collapse of the inner core of a stratified
cloud takes place.
Here we have $\hat\rho_0=1$ and $\hat\rho=\psi^2$, so that
the complete solution is given by
\begin{eqnarray}
\rhohat & = & \frac{4}{\left(2-\tau\right)^2}, \label{eq:h1} \\
\rhat^3      & = & \frac{1}{4}m\left(2-\tau\right)^2, \label{eq:h2} \\
\vhat_n & = & 
              \frac{\left(2m\right)^{1/3}}{\left(2-\tau\right)^{1/3}},
                \label{eq:h3} \\
b & = & \Biggl[ 1 + \frac{3}{2}
    \frac{2^{8/3}\,\delta}{\left(2-\tau\right)^{8/3}}\biggl(1-m^{2/3}\biggr) 
     \Biggr]^{1/2}, \label{eq:h4}
\end{eqnarray}
and
\begin{eqnarray}
\phi & = &
\frac{4}{9}\,\frac{1}{\delta}\,\frac{\left(2-\tau\right)^4}{16}
\Biggl\{\Biggl[ 1 + \frac{3}{2}
    \frac{2^{8/3}\,\delta}{\left(2-\tau\right)^{8/3}}\Biggr]^{3/2} -
\nonumber \\
      & & \Biggl[ 1 + \frac{3}{2}
    \frac{2^{8/3}\,\delta}{\left(2-\tau\right)^{8/3}}\biggl(1-m^{2/3}\biggr) 
     \Biggr]^{3/2} \Biggr\}. 
\label{eq:h5}
\end{eqnarray}

        There are several things that are worth noting. First, $\rhohat$
{\it remains uniform\/} even at late times, just as in the collapse of a
non-magnetic spherical cloud (Spitzer, 1978).
The collapse is homologous, with $v\propto r$, since the acceleration
rises linearly with $r$ for a homogeneous cloud.
Figure~1 is a plot of
$\rhat$ as a function of $\tau$ for various values of $m$,
in a  cloud with $\delta=4\times 10^{-3}$, which is appropriate
for a cloud of $\mcloud=5\,\msol$, $R_0 = 0.45\pc$, and $B_0 =
30\,\mu\gauss$; 
such a cloud has $\Mphio/M_0=8.3$, which is
quite magnetically subcritical. 
Next, note that the  expressions for the magnetic field
and the flux introduce a characteristic time,
$\taucr$, which we choose to define as
\begin{equation}
2-\taucr=2(3\delta )^{3/8}.
\label{eq:taucr}
\end{equation}
For the case at hand, $\taucr=1.62$.

        Because the outer
parts of the cloud accelerate first, one is tempted to 
conclude that it is the cloud as a whole that first
becomes  gravitationally
unstable, rather than the central region. 
This hypothesis can be tested by evaluating $\Mphi/M$ as a function
of time and mass.
For flattened clouds, it is customary to evaluate this ratio
for a flux tube.  In our spherical geometry, however, we 
evaluate $M(r)$, the mass inside a sphere of radius $r$, and compare this
with the flux $\Phi(r)$.  Had we included the pressure that makes
the cloud spherical, the condition for gravitational instability 
would have been that $M(r)$ exceed both the Jeans mass and the
magnetic critical mass $\Mphi$.  Having $M(r)>\Mphi(r)$ 
is thus a {\it necessary} condition for gravitational instability.
Figures 2a and 2b show
\begin{equation}
\frac{\Mphi}{M}=\frac{\Mphio}{M_0}\left(\frac{\phi}{m}\right)
        =\frac{0.53}{\delta^{1/2}}\left(\frac{\phi}{m}\right),
\end{equation}
where we have made use of equation (\ref{eq:critratio0}).  
Note that at $\tau=0$ the cloud is magnetically subcritical
everywhere ($\Mphi/M > 1$), 
and that the outer edge of the cloud
first becomes unstable at $\tau\simeq 1.62= \taucr$. 
More precisely, at $\tau=\taucr$, we have $\phi(m=1)=1.93\delta^{1/2}$,
so that $\Mphi/M_0=1.03$, independent of the value of $\delta$.
Thus, $\taucr$ is very nearly equal to the time at which the cloud
first becomes supercritical.

        These results 
show that in a homogeneous, cold, and magnetized cloud the collapse
due to ambipolar diffusion proceeds from {\it outside-in}
  \footnote{It is
  interesting to note that this outside-in collapse has been found by
  Goldreich \& Weber (1980\markcite{gow80}) in a completely different
  context. They studied the core collapse in supernova progenitors
  with an equation of state $P\propto\rho^{4/3}$. In our case, this
  equation of state becomes appropriate once the magnetic field has
  been amplified considerably, so that $b\gg 1$ (see eqs.~\refeqp{h1}
  and \refeqp{h4}).
  Goldreich \& Weber's results show that
  the dynamical evolution of the core is homologous, with the outside of the
  core accelerating first.}.
The inner regions of the cloud are gravitationally stable, and
collapse only because of the weight of the overlying material.  
By contrast, the collapse of a non-magnetized singular isothermal
sphere occurs inside-out (Shu 1977): such a sphere is unstable at all
points, but the collapse begins at the origin, where the free-fall time
is the shortest.
        
        Finally, we determine $\phi_{\rm core}$,
the residual flux trapped in the cloud when
it collapses to a singularity at $\tau=2$.  Taking the limit of
$\phi(m=1)$ as $\tau\rightarrow 2$ in equation (\ref{eq:h5}), we find
$\phi_{\rm core}=(2\delta/3)^{1/2}$.  As a result, we have
\begin{equation}
\frac{M_{\Phi, {\rm core}}}{M_0}=
        \left(\frac{M_{\Phi 0}}{M_0}\right)\phi_{\rm core}=0.43.
\label{eq:trapped}
\end{equation}
Our model thus
shows that the trapped flux is sufficiently small that the collapsed
core is magnetically supercritical, as expected.

        Are these results consistent with our assumptions? 
The validity of the quasi-static approximation
can be readily assessed with the help of the parameter $\alpha$---the
acceleration of the neutrals in units of the local gravitational
acceleration (eq. \refeqp{alphadef})---and  the expressions in 
eqs. \refeqm{h2} and \refeqm{h3}, which yields
\begin{equation}
\frac{\rhat^2}{m}\dtaufixm{\vhat_n} = \frac{1}{6}. \label{eq:h6}
\end{equation}
Therefore
\begin{equation}
\alpha  = 2.1 \times 10^{-3}, \label{eq:h7}
\end{equation}
and the approximation of quasi-static evolution is very good, at least
so long as our static ion approximation holds.

        Next, the assumption of negligible ion velocity compared to the
neutral velocity has to be checked. Figure 3 is a plot of $v_i/v_n$ as
a function of the Lagrangian coordinate $M$ (in units of the cloud's
mass $\mcloud$) for four different times. It can be seen that up to
$t/\tad = 1.62$---i.e., at $\taucr$, 
when the cloud as a whole becomes magnetically
supercritical--- $v_i/v_n \la 0.3$; 
therefore our assumption
of negligible $v_i$ is justified {\it a posteriori}, up to the
point where the cloud becomes fully supercritical. Once the cloud is
supercritical, $v_i$ begins to approach $v_n$, and our solution breaks
down. This accounts for the late-time behavior of $\Mphi/\mcloud$ seen
in Figure 2b: after $\tau=1.62$ this ratio does not change much, as
expected when the magnetic flux starts to become frozen onto the
neutrals.
We conclude that the static ion approximation 
breaks down at about the same time the cloud becomes magnetically 
supercritical; this
ultimately leads to the breakdown of the quasi-static approximation
for the neutrals.

        We can analytically 
verify that the static ion approximation breaks down
when the cloud becomes supercritical.  Note that
the maximum value of $v_i/v_n$ occurs at the center of the cloud.
According to equation (\ref{eq:vivncomp}), $v_i/v_n$ depends on
the ratio of $\partial\phi/\partial\tau$ to $\partial\phi/
\partial\rhat$.  Near the center of the cloud, we have
$\phi\simeq \pi\rhat^2b$ ($m\ll m_c$).  Evaluating the partial derivatives,
we find
\begin{equation}
\left(\frac{v_i}{v_n}\right)_{m=0}=\left[1+2\left(\frac{2-\tau}{2-\taucr}
        \right)^{8/3}\right]^{-1}.
\end{equation}
This result confirms that the ions are approximately static prior to
$\taucr$, when $v_i/v_n=1/3$.
Note that this estimate indicates that the ions and neutrals become
comoving as $\tau\rightarrow 2$, as they do in reality.

        A remarkable feature of the solution discussed in this
section is that it is completely independent of one of the
basic parameters of the problem, 
the ratio of the ambipolar diffusion time to the free
fall time ($\nuff \propto \tad/\tff$).
The reason for this is now clear: our solution is valid only
during the magnetically subcritical, quasi-static 
stage of evolution, when the timescale
is determined by $\tad$ alone.  The solution sets the initial
conditions for the subsequent supercritical, dynamical stage
of evolution, when the timescale is set by $\tff$.

\section{A Stratified, Non-Singular Cloud}

We now consider a more realistic case, namely, that of a stratified,
non-singular cloud. For an initial density profile of the form
\begin{equation}
\function{\rhohat}{\tau=0,m} = \frac{\mc^2}{\left(\mc + m\right)^2} 
\label{eq:p1}
\end{equation}
(i.e., choosing $p=2$ in eq. \refeqp{42}) equation \refeqm{3.6} can be
inverted algebraically, and we obtain
\begin{equation}
\rhohat = \frac{4}
{\biggl[ 1 + \frac{m}{m_c} - \tau + \sqrt{\left(1+\frac{m}{m_c}\right)^2\,
- 2\tau\frac{m}{m_c}}\biggr]^2}.\label{eq:p2}
\end{equation}
Recall that this case is a good approximation to a stable,
 non-singular, isothermal sphere,
which is relevant to the study of real clouds.

        Armed with the solution in equation \refeqm{p2},
we can solve for $\function{\rhat}{\tau,m}$ using equation 
\refeqm{27} and the boundary conditions\footnote{The boundary
  condition given by equation~\refeqm{rhatbc2} is strictly valid for $\tau
  <2$ because $(\tau=2,m\le m_c)$ is a singular point of
  equation~\refeqm{27}. However, the solution for $\rhat(\tau,m)$ given by
  equation~\refeqm{p3} can still be used for $\tau >2$ and $m>m_c$. The
  reason is that the evolution of shells enclosing  $m>m_c$ depends,
  by Newton's theorem, on
  $m$ only and not its distribution---whether pointlike or distributed
  over a finite volume. Therefore, the neutral's velocity $\vhat_n$ and
  the location of each shell $\rhat$ are smooth
  functions of $\tau$ across $\tau=2$ for $m>m_c$.}
in equations~\refeqm{rhatbc1}-\refeqm{rhatbc2}:
\begin{eqnarray}
\function{\rhat^3}{\tau,m} = \frac{\mc\,\epsilon}{4}&\Biggl\{ 
        \displaystyle{\frac23\left[\left(1+\frac{m}{m_c}\right)^2-
        2\left(\frac{m}{m_c}\right)\tau\right]^{3/2}
        +\frac23\left(1+\frac{m}{m_c}\right)^3} \nonumber\\
        &\displaystyle{+\left(\frac{m}{m_c}\right)\tau^2-
        2\left(\frac{m}{m_c}\right)\left(1+\frac{m}{m_c}\right)\tau-
        \frac43\Biggr\}}.
\label{eq:p3}
\end{eqnarray}

        What is the nature of the solution given by equations \refeqm{p2} and
\refeqm{p3}? Recall that  at
the time $\tau=2$, a core of mass $M_c=\mc\,\mcloud$ collapses 
into the origin,
leaving behind an envelope with mass $(1-\mc)\mcloud$.
Evaluating $\rhohat$
and $\rhat^3$ at $\tau=2$, they take the form
\begin{equation}
\function{\rhohat}{\tau=2,m} = 4\left[ \frac{m}{\mc} - 1 + 
                              \left|1-\frac{m}{\mc}\right|\right]^{-2}
\label{eq:p5} 
\end{equation}        
and
\begin{equation}
\function{\rhat^3}{\tau=2,m} = \frac{\mc\epsilon}{6} 
               \biggl[ \left(\frac{m}{\mc}-1\right)^3 + 
                       \left|\frac{m}{\mc}-1\right|^3 \biggr], \label{eq:p6}
\end{equation}
respectively. Note that for $m \le \mc$ 
the expression between the square brackets  in equation~\refeqm{p6} 
vanishes, and $\rhat=0$ identically.
We conclude that in a cloud
that is initially everywhere magnetically subcritical, a core with
a fraction $\mc$ of the total cloud mass will, through ambipolar
diffusion, eventually become supercritical, and collapse into the
origin. The relationship between $\mc$ and the initial degree of
concentration of the cloud, $\epsilon$, given by equation \refeqm{3.4} with
$p=2$, is
plotted in Figure 4. We find that $\mc=1$ for $\epsilon=3/7$, and
therefore clouds with $\epsilon \ge 3/7$ will collapse as a whole,
leaving no envelope behind. 

\subsection{Evolution Prior to Collapse}

Further insight into the nature of the solution can be gained by
examining in detail the time evolution. Here we consider the evolution
up to the formation of a singularity in the origin
of a cloud with $\epsilon = 0.1$ (corresponding to $\mc=0.26$)
and $\delta=4\times 10^{-3}$ (or,
equivalently, $M_{\Phi,0}/\mcloud = 8.4$). These
values correspond to a  cloud of mass
$\mcloud=5\,\msol$, initial radius $R_0=0.45\pc$, a magnetic field
strength at the edge of the cloud $B_0=30\,\mu\gauss$, and an initial
particle density at the center of $n_{c,0}=2.2\times 10^3\,{\rm
  cm^{-3}}$---yielding a characteristic ambipolar diffusion time
$\tad=3.30\,{\rm Myr}$.

Figure 5 is a plot of the 
Lagrangian position $r$, neutral velocity $v_n$, density
$\rho$, and  magnetic field strength $B$.
Focusing on Figure 5a, it is noteworthy that by $\tau=1.86$ 
the radius for $m=1$ is reduced by a factor $\sim 0.6$
from its initial value , while for
$m\la\mc$ this change is by a factor of order $\sim 0.2$.
Moreover, note how the 
change in radius is roughly the same for all $m$ in the region
$m\la\mc$,  i.e, the
evolution of the core resembles that of a homogeneous cloud. This
result is further illuminated by considering the behavior of
$v_n$. Initially the absolute value of $v_n$ is largest at
the outer edge of the cloud, but by $t/\tad \sim 1.6$ a maximum (in
absolute value) starts to develop at $m\sim\mc$, which becomes more
accentuated as time goes on. In other words, the dynamical behavior of
the core with $m\la\mc$ resembles that of a homogeneous cloud, and its
collapse also proceeds from {\it outside-in}.

Finally, it is interesting to examine Figure 5b and note that by
$t/\tad=1.8$ the central density has increased by a factor $\sim 100$,
while $B$ at the center has 
increased by only a factor $\la 4$---as expected in a
cloud whose evolution is dominated by ambipolar diffusion.

\subsection{Self-Consistency}

In this section we check the self-consistency of our results for our
fiducial stratified cloud, namely, the validity of the quasi-static
approximation, the neglect of thermal stresses, and the assumption
that $v_i \ll v_n$, which allows us to neglect the induction equation.

\subsubsection{Quasi-static Evolution}  

Figure 6 is a graph of $\alpha$,
 the acceleration of the neutrals
in units of the local gravitational acceleration, as a
function of $m$ and $\tau$. 
Note that, up to the time shown ($\tau = 1.996$),
$\alpha \ll 1$,  i.e., the acceleration of the neutrals is a very
 small fraction of the local gravitational acceleration,
and therefore the assumption of quasi-static evolution
is very good. Also, note that for $m \la 0.2$, $\alpha$ is roughly
constant, as in a homogeneous cloud.
However, just as in the case of the homogeneous cloud, it is 
the breakdown of the static ion approximation that occurs first,
and that ultimately leads to dynamical evolution.

\subsubsection{Static-Ion Approximation}
            
        A key assumption in our approach is that the ion velocity is much
smaller than the neutral velocity, so that the
hydrodynamic and magnetic problems are decoupled.  This
key assumption has to be verified {\it a posteriori} by computing the
magnetic flux $\Phi$ (eq. \refeqp{fluxdef}) and computing $v_i$ as
prescribed by equation \refeqm{vicheck}.
The evolution of the magnetic flux as a function of mass and
of position is shown in Fugure 7.
Figure 8 is a plot of $v_i/v_n$ as a function of the dimensionless
Lagrangian coordinate $M/\mcloud$ for five different times 
(in units of the characteristic ambipolar diffusion
time \tad). Note that for $\tau \la 1$,
$v_i/v_n \ll 1$ obtains everywhere in the cloud, justifying our initial
assumption.  By $\tau=1.64$,   $v_i/v_n \approx 0.35$ in the inner
few percent (by mass) of the cloud.  This result is similar to the 
one we obtained for the homogeneous cloud above.
By plotting $\Mphi/M$ (Figure 9),
we see that a part of the cloud first becomes magnetically
supercritical shortly before $\tau=1.64$.  Hence, just as
in the homogeneous case, the static ion approximation breaks
down when the cloud becomes supercritical (although in
this case only the core region is supercritical).

\subsubsection{The Validity of the $T=0$ Approximation}

        One of our main simplifying assumptions is that thermal stresses are
negligible compared to magnetic stresses. The validity of this
approximation can
be checked {\it a posteriori} by looking at the value of
$\beta^\prime$, 
i.e, the ratio of the thermal to magnetic pressure gradients 
(eq. \refeqp{betaprime}).
Figure 10 is a plot of  $\beta^\prime$ as a function of $m$ and $\tau$
for our fiducial cloud. Note that $\beta^\prime \ll 1$ throughout the
evolution; therefore, our assumption that thermal stresses can be
neglected is justified. 

        At a given time (e.g., for $t/\tad=0$),
our model predicts that
 the importance of thermal support would increase
as one moves towards the center of the cloud 
until 
$m\la\mc$, where it
starts to decrease. In a real cloud, one expects that, at the center,
thermal pressure becomes important for a  mass of order the Jean mass at the
appropriate 
central density. Thus, one expects $\beta^\prime$ to
increase towards the center of the cloud; whether at the center 
$\beta^\prime\sim 1$
depends on the particular magnetic configuration. Also, as
 ambipolar diffusion proceeds, the central density increases, and the
 mass---and size---of the region where thermal stresses are important
 {\it decreases} with time 
(recall that the Jeans mass $\propto (n_c)^{-1/2}$,
 where $n_c$ is the central density). In our case, the local maximum of
 $\beta^\prime$ obtains at a fixed value of $m$---
namely,
$m=\mc$---because of
 our choice of initial conditions and our neglect of thermal
 stresses. However, this is not critical as long as we find that
 $\beta^\prime\ll 1$ throughout the cloud, because, as mentioned
 above, the Jeans mass decreases with increasing density. Thus, at
 worst, our calculation is in error in a region that becomes
 increasingly smaller with time. 

        As time increases,
thermal pressure forces at the center
{\it may\/} begin to dominate the magnetic stress 
due to the loss of magnetic flux. The
answer to this question, again, depends on the particular magnetic
configuration. 
In our case, we find that thermal stresses never become
important during the quasistatic phase; this finding is in agreement
with the results of Fiedler \& Mouschovias (1992), who find that even during
the dynamical collapse of the core, magnetic stresses dominate over
thermal pressure
at least up to a density of $\sim 10^9$ cm$^{-3}$.
In our model,
if there is no core to begin with--the
constant density case-- one never forms;
(on the other hand, 
cores can form during the
subsonic evolution of clouds supported by gas pressure---see
Bodenheimer \& Sweigert, 1968\markcite{bs68}). Once a core has formed,
then we find that
the subsequent contraction of the cloud results in a {\it decrease\/}
in the relative importance of thermal stresses,
as shown 
in Figure 10.

Note that this predominance of magnetic support over thermal support
does not contradict the observational evidence that the linewidths of
dense cores are essentially thermal (see, e.g., Caselli \& Myers,
1995\markcite{cam95}, and references therein).
Recall that, in 
our
 picture, the
support of the cloud is due to {\it quasistatic\/} magnetic
fields---as opposed to magnetic turbulence---and that the drift
velocities during the quasistatic phase are an order of magnitude
below the speed of sound (see Figure 5a). Therefore, in 
our model,
the main contribution to the line widths is thermal motion.

\subsection{Comparison with the Numerical Work of Fiedler \& Mouschovias}

Are our results relevant to the evolution of real clouds? To date, one 
can obtain an answer to this question only by comparison with
state-of-the-art numerical calculations. Here we compare our results
against those of Fiedler \& Mouschovias (1992)\markcite{fm92}
for their baseline model, Model 1.
To effect the comparison, we need to
specify five parameters: the time $t_0$ in Fiedler \& Mouschovias's 
results that we identify with $\tau=0$, the corresponding central
density \rhoc, the mass and initial radius of the cloud, \mcloud\ and
\rzero\ respectively, and the magnetic field at the edge of the cloud, \bzero.
The values of these last three parameters are fixed by Fiedler \&
Mouschovias's  choices, namely $\mcloud = 45.5\,\msol$, $\rzero =
0.75\,{\rm pc}$, and $\bzero=30\,\mu{\rm G}$; the values of  $t_0$ and \rhoc\
require further consideration.

Because Fiedler \& Mouschovias start their calculations with a
homogeneous, uniformly magnetized cloud that relaxes to a quasi-static
state, the only practical way to determine $t_0$ and \rhoc\ is by
fitting their results for the central
density during the quasi-static phase with
our expression for the central density (eq.~\refeqp{3.7} in dimensional form):
\begin{equation}
\rho_{{\rm c}}\,(t) = 
4\,\rhoc\biggl[2-\frac{4\pi G \rho_{{\rm
c},0}^{1/2}\,(t-t_0)}{\kni}\biggr]^{-2} \label{eq:p2fit}.
\end{equation}
Using the values adopted by Fielder \& Mouschovias in their Model 1 
for the mean
molecular weight, $K_i$, and $\left<\sigma v\right>_{ni}$, we find 
that there is a continuum of
equally good fits with $n_{{\rm c},0}\ga 1.5\times 10^3\,{\rm
  cm^{-3}}$---or, equivalently, $\epsilon \la 0.3$---and 
$t_0 \ga 8.0\,{\rm Myr}$. However, we show below that geometrical
constraints narrow the range to $\epsilon \la 0.1$ ($n_{{\rm c},0}\ga
4.5\times 10^3\,{\rm cm^{-3}}$) and $t_0 \ga 10.0\,{\rm Myr}$. We
decide to adopt our fiducial value of $\epsilon=0.1$, and a very good
fit obtains with $t_0=11.85\,{\rm Myr}$. Given that \mcloud\ and
\rzero\ are fixed, this choice of $\epsilon$ translates into an initial
central density $n_{{\rm c},0}=4.4\times 10^3\,{\rm cm^{-3}}$, and
the characteristic
ambipolar diffusion time for the cloud is $\tad=2.33\,{\rm
  Myr}$. Finally, the choice $\bzero=30\,\mu\gauss$ translates into 
$\delta=4.1\times 10^{-2}$---a factor of
ten larger than our fiducial value---with $M_{\Phi,0}/\mcloud=2.6$.

        Figure 11 is a comparison of the evolution of  the central density 
and central magnetic field---this last 
obtained from evaluating $B$ at $m=10^{-3}$---between
our model and Fiedler \& Mouschovias's. Focusing first on the central
density, we see that even though $t_0 =
11.85\,{\rm Myr}$, the fit is good starting at $t\approx 8\,{\rm
Myr}$ and remains good until $t\approx 16\,{\rm Myr}$ 
($t/\tad \approx 1.8$).
After $t=16\,\myr$ the two models diverge,
and no fit can be found for any choice of $t_0$ and $n_{{\rm c},0}$.
This is due to the onset of dynamical collapse in the Fiedler \&
Mouschovias calculation, while our model still evolves
quasi-statically up to $t\sim 16.5\,\myr$ (refer back to Fig. 6). The
reason for this discrepancy can be found in the evolution of $B_c$ in
Figure 11: note how the strength of the central magnetic field
increases faster in our model than in Fiedler \& Mouschovias'---by
$t\approx 15\,\myr$ the difference is a factor $\sim 2$, and
thereafter we find that $B_c \propto n_c^{2/3}$, i.e. flux-freezing
obtains in the core.
This larger
magnetic field provides enhanced support against gravity and the onset
of dynamical collapse is retarded. However
it is interesting to note that at $n_c \sim
10^5\, {\rm cm^{-3}}$, where our solution starts
to differ significantly from that of Fiedler \& Mouschovias, the
 free-fall time in the core is $\sim 0.1\
\myr$ and the ambipolar diffusion time is $\sim 0.5\,\myr$. Therefore,
our solution for $n_c$ agrees with that of Fiedler \& Mouschovias up to the
point where the evolution of the core becomes dynamical and the core is within
{\it a few free-fall times} from reaching the origin, {\it
  notwithstanding the large discrepancy in $B_c$.} 
This is so even though the static ion approximation
breaks down  earlier than in our fiducial cloud---we find that in the core 
$v_i/v_n \sim 1$ by $\tau\sim 1.4$ due to the fact that
$\delta$ is an order of magnitude larger than the
fiducial value used in our numerical examples.

        So far we have only proven that our solution compares well
with that of Fiedler \& Mouschovias in the central region. 
Given the discrepancy in the late-time evolution, it is
interesting to compare the evolution in other parts of the cloud.
Because of the differences between the two calculations, the only
meaningful comparison is that between our results and those of Fiedler
\& Mouschovias for the {\it midplane} density of their cloud; this
comparison is presented in Figure  12, which is a plot of density as a
function of radius for three different times,
$t=10.2\,\myr$, $t= 15.1\, \myr$, and $t=16.0\,\myr$.
Two things are noteworthy in this figure. First, even though the
central density disagrees by a factor of $\sim 2$, the
overall {\it profile\/} is well reproduced\footnote{However, note that
  for $t=16.0\,\myr$, where the central density is reproduced most
  accurately, the drop in density at the edge of the core is steeper
  in the Fiedler \& Mouschovias model than in ours. This is a natural
  result of neglecting thermal pressure: the density contrast between the
  center and the edge of a {\it stable}, thermally supported 
  cloud is 
$14:1$ (Bonnor, 1956\markcite{bon56}), and the mass of
  the core decreases with increasing density as $\rho^{-1/2}$; in our
  model the mass of the core is fixed, while the density contrast
  between the center and the edge of the core increases with time from
  $4:1$ at $\tau=0$ (eq. \refeqp{p2}).}.
Second, note how the density
profile in the envelope is also well reproduced ($\rho \propto r^{-2}$), but
Fiedler \& Mouschovias's values are consistently larger than ours
(again, by $\sim 2$); this is a natural consequence of the different
geometries. Although Fiedler \& Mouschovias's model is fairly
spherical---with an aspect ratio $1$:$3$ in the envelope---our
cloud is strictly spherical, so our density in the outer parts has to be
lower for the two clouds to have the same total mass. This constraint
guided us in choosing $\epsilon=0.1$ for the comparison. Otherwise, for
$\epsilon > 0.1$, the spatial extent of our core  and our 
envelope density would be
{\it larger\/} than those in Fiedler
\& Mouschovias's cloud, and the clouds' masses would not match.

        These results confirm that the assumptions of negligible
thermal stresses
and of static ions
have minimal consequences for our results for the hydrodynamical
variables until the latest stages of gravitational collapse.

\subsection{ The Initial Protostellar Mass}

What fraction of the cloud becomes the actual protostar?
In our model, the answer is
 $\mc\mcloud$, which is fixed {\it ab initio} once the initial
degree of concentration of the cloud $\epsilon$ is fixed. This
characteristic mass corresponds to the inner fraction of the cloud
that first becomes magnetically supercritical and collapses, detaching
itself from the remaining envelope. However, \mc\ is a free parameter
in our model; what determines in nature, then, the  protostellar
mass? Notwithstanding the shortcomings of our model, here we show that
it provides insight to this question.

        As we noted before, 
the stability of a cold cloud is governed by the ratio
$\Mphi/M$, where \Mphi\ is the magnetic critical mass
(eq. \refeqp{mphidef}). Those regions of the cloud with $M >
\Mphi$ are dynamically unstable and will collapse even if the magnetic
field is perfectly coupled to the neutral material. Figure 9 is a plot
of $\Mphi/M$ as a function of time and the Lagrangian coordinate $M$
for our fiducial cloud ($\epsilon=0.1$ and
$\delta=4\times10^{-3}$, with $M_{\Phi,0}/\mcloud=8.4$). 
Two things are noteworthy in this figure:
first, the cloud is initially magnetically subcritical throughout, and
as time increases the region $M\sim M_c$ first becomes
supercritical
at $t/\tad \sim 1.5$. Second, and more important, at any given time 
the ratio $\Mphi/M$ has a global
minimum as a function of $M$, and this minimum occurs at {\it finite\/}
$M$. Thus, a region of finite mass first becomes magnetically
supercritical, and the collapse proceeds inward from this region. This
result is readily understood: for small enough values of $m$, both $B$
and $\rho$ are fairly constant, so that $\Mphi \propto BR^2 \propto
M^{2/3}$. On the other hand, in the outer parts of the cloud
$\rho\propto r^{-2}$ and $B\sim$constant, so that $\Mphi\propto R^2
\propto M^2$. Thus altogether $\Mphi/M \propto M^{-1/3}\, + K M$,
where $K$ is a constant, and this expression has a minimum at a mass
at which the two terms are comparable.

        Is this result relevant to clouds with a finite temperature? The
answer is yes---as long as magnetic stresses dominate outside the
thermally supported core.
In a real
cloud, there are {\it two} characteristic masses, namely a Jeans mass
which can be
supported by thermal stresses only, $M_{\rm J}$, and the magnetic
critical mass \Mphi; and both are time dependent. Therefore, at any time 
the stability of the cloud is now determined by the ratio
$(\Mphi + M_{\rm J})/M$ (McKee, 1989\markcite{mck89}).
If the core is thermally supported, this function
will also have
a global minimum as a function
of $M$. Just as before, this minimum will occur at a mass at which the
two terms are comparable---i.e., $\Mphi \sim M_{\rm J}$---so that
$M$ is of order $M_{\rm J}$. We conclude that the characteristic protostellar
mass is of order the Jean mass at the onset of dynamical evolution
(see also Larson 1985\markcite{lar85}, 1995\markcite{lar95}):
\begin{eqnarray}
  \label{eq:mjeans}
  M_{\rm J} &=& 1.18\,
                \frac{c_s^4}{\left(G^3\,P_0\right)^{1/2}}\nonumber \\
            &=&
                3.6\,\left(\frac{T}{10\,\kelvin}\right)^2\,
                     \left(\frac{P_0/k}{10^4\;\kelvin\;\ccc}
                      \right)^{-1/2}\;\msol,
\end{eqnarray}
where $c_s$ is the speed of sound and $P_0$ is the thermal pressure
confining the cloud,
which is typically somewhat greater than the thermal pressure in
the diffuse ISM (Bertoldi \& McKee 1992\markcite{bm92}).  Note that if the
core is thermally supported, then the thermal pressure is
comparable to the magnetic pressure and 
the characteristic stellar mass $M_J\propto 1/B$
(cf. Basu \& Mouschovias 1995b\markcite{bm95b}).  In our model, the role of the
Jeans mass is taken by the core mass $M_c$.

\subsection{The Protostellar Magnetic Flux}

        Our model allows us to estimate the 
initial magnetic flux trapped in the protostellar core,
$\Phi_{\rm core}(\tau=2)$.
We have already found the trapped flux analytically for the
homogeneous case; here we determine it for the $p=2$ case.
We can do this despite the breakdown of the static ion
approximation before $\tau=2$ because the breakdown of
the approximation leads to $v_i\sim v_n$, so that 
flux freezing obtains.  As a result, the value of
$\Phi_{\rm core}(\tau=2)$ is essentially the value of 
$\Phi_{\rm core}$ at the time the approximation breaks down.
This is shown in Figure 13,
which is a plot of the evolution of the magnetic flux in the core,
$\Phi_{\rm core}$, 
in units of the cloud's initial magnetic flux,
$\Phi_0$, for $\delta=10^{-3}$ and different values of
$\epsilon$. Note that the amount of flux trapped in the core at
$\tau=2$ is a decreasing function of $\epsilon$, or, equivalently, of $\mc$.

        How does $\Phi_{\rm core}(\tau=2)/\Phi_0$ scale with $\mc$ and 
$\delta$? We can
answer this question by examining equation~\refeqm{28}. To isolate the
dependence on $\mc$, let us introduce a new Lagrangian variable
$y=m/\mc$. Then, equation~\refeqm{28} reads
\begin{equation}
  \label{eq:phic1}
  \left.\frac{\partial b^2}{\partial y}\right|_\tau = -\delta
  \mc^2\,\frac{y}{\rhat^4}. 
\end{equation}
According to equation~\refeqm{p3}, we can write $\rhat^4 =
(\mc\,\epsilon)^{4/3}\,f(\tau,y)$.
Let us assume $m_c\ll 1$, so that $\epsilon\simeq 3m_c^2$
(eq.~\refeqp{3.4} with $p=2$). We then
obtain from equation~\refeqm{phic1}
\begin{equation}
  \label{eq:phic2}
  b^2(\tau,y) = 1 + \frac{3\delta}{ m_c^2}\,\int_y^{1/m_c}\,
        \frac{y\,dy}{f(\tau,y)}.
\end{equation}
In the core ($y\le 1$), and for $\tau>1$, 
we have $b \gg 1$ (see Figure 5b),
and therefore the first term on the r.h.s of equation~\refeqm{phic2} can be
neglected. Furthermore, because the gradient of $b$ is largest just
outside the core (again, see Figure 5b), the integral in
equation~\refeqm{phic2} depends weakly on the upper limit of integration,
so that
\begin{equation}
  \label{eq:phic3}
  b(\tau>1, y\le 1) \propto \delta^{1/2}\,m_c^{-1}.
\end{equation}

As for the magnetic flux in the core,  by 
rewriting equation~\refeqm{fluxdef} in Lagrangian coordinates 
and integrating we obtain, in units of
$\Phi_0$,
\begin{equation}
  \label{eq:phic4}
 \phi_{\rm core} = \frac23\epsilon\mc\,\int_0^1\,\frac{b\,dy}{\rhat\,\rhohat}.
\end{equation}
Because $\rhohat$ depends on $\tau$ and $y$ only (eq.~\refeqp{p2})
whereas $\rhat$ scales as $(m_c\epsilon)^{1/3}$, we
finally obtain that at late times
\begin{equation}
  \label{eq:phic5}
  \phi_{\rm core}(\tau >1) \propto \mc\,\delta^{1/2}.
\end{equation}
The scaling in equation~\refeqm{phic5} describes our numerical
results quite well, 
and we find the coefficient of proportionality to be $1.16$.
By analogy with equation (\ref{eq:trapped}), we find that 
the value of $\Mphi/M$ in the core is
\begin{equation}
\frac{M_{\Phi, {\rm core}}}{M_c}=
        \left(\frac{M_{\Phi 0}}{M_0}\right)\frac{\phi_{\rm core}}
        {m_c}=0.61,
\label{eq:trapped2}
\end{equation}
independent of $m_c$ or $\delta$.  Just as in the case of the
homogeneous cloud, the flux is reduced until the core is
supercritical; the more subcritical the core is to begin with, the
greater the flux lost.

\subsection{Post-Collapse Evolution}

A particularly attractive feature of our solution is that we can follow,
with the help of equation \refeqm{p2},
the evolution of the envelope with $m>m_c$ after the collapse of the core
at $\tau=2$. 

As in the pre-collapse evolution, one should check for
self-consistency regarding our quasi-static approximation, the neglect
of thermal stresses, and the assumption that $v_i\ll v_n$. The
validity of the quasi-static approximation is assessed with the
help of $\alpha$ (eq. \refeqp{alphadef}), the local ratio of the
neutral's acceleration to the gravitational acceleration; similarly,
the validity of the $T=0$ approximation is verified by
evaluating $\beta^\prime$ (eq. \refeqp{betaprimecomp}). On the other
hand, the assumption of negligible motion of the ions requires special
consideration.

Verification of the assumption $v_i/v_n \ll 1$  requires $\Phi(t,r)$
to be known. It is difficult to evaluate $\Phi(t,r)$
after the collapse of the core in the same way as
for $\tau <2$ (integrating eq.~\refeqp{fluxdef} outwards from $r=0$)
because, as more and more mass shells accrete onto the
protostar, the protostellar magnetic flux increases with time, and its
evaluation requires the solution of a partial differential equation.
However, one can evaluate the magnetic-flux loss at the edge of the
cloud and find $\Phi(M=\mcloud)$ as a function of time  in 
the following way. Adding and
substracting $v_n\,\partial\Phi/\partial r$  
to the l.h.s of equation~\refeqm{fluxfreeze1},
substituting $\partial\Phi/\partial r$ from
equation~\refeqm{fluxdef}, and 
switching to Lagrangian coordinates we obtain
\begin{equation}
  \label{eq:acflux1}
  \dtfixm{\Phi} = 2\pi v_n\,\left(1-\frac{v_i}{v_n}\right)\,Br
        \simeq 2\pi v_nBr,
\end{equation}
where the last step follows for $v_i\ll v_n$.
Note that at the edge of the cloud we have $B=B_0$, so 
that equation~\refeqm{acflux1} may be integrated to evolve
the magnetic flux at $M=\mcloud$.
With this value at hand,
we can then integrate equation~\refeqm{fluxdef} {\it inwards}.
It turns out that integrating inwards
equation~\refeqm{fluxdef} with the initial value found from
equation~\refeqm{acflux1} starts 
to breakdown at the point where $v_i\sim v_n$ (we obtain negative
values for $\Phi$; recall that $v_n$ is negative definite), 
but, as long as $\alpha\ll 1$, the assumption
$v_i\ll v_n$ is very good throughout the quasi-static
envelope\footnote{The same method may be applied to find $\Phi$ during
  the pre-collapse stage. We find that as long as $v_i \ll v_n$ the two
  methods (integrating  outwards from $r=0$ and integrating
  inwards from $M=\mcloud$) agree very well. However, as soon as $v_i$
  approaches $v_n$ in the core, the integration from $M=\mcloud$ breaks down
  for $m\le m_c$, although it
  yields $v_i \ll v_n$ throughout the envelope.}.
Finally, we find that the assumption of negligible thermal stresses is very
good throughout the post-collapse evolution and that our results are
only limited by the breakdown of the quasi-static approximation. We
arbitrarily deem the evolution to become dynamical where $\alpha\geq 0.1$.

        Figure 14 is a plot of the neutral velocity and density in our
fiducial cloud as a function
of position $r$ for six different times after the collapse of the
core; for each curve, the smallest value of $r$ is that where
$\alpha=0.1$, while the largest corresponds to the edge of the cloud.
For $\mcloud=5\,\msol$ and $\rzero=0.45\pc$---and recalling that
$\epsilon=0.1$---we obtain  $n_{c,0}= 2200\,\ccc$ and $v_{n0}=4.5\times
10^{-3}\,{\rm km\,s^{-1}}$; therefore the typical densities in the
envelope are $\ga 750\,\ccc$ and the maximum velocities are $\sim
-0.10\,{\rm km\,s^{-1}}$---a factor $\la 2$ smaller than the speed of sound for
$T=10\,\kelvin$. 

Two things are noteworthy in Figure 14. First, after the collapse of
the core at $t/\tad=2.0$ the density profile starts to flatten as the
velocity profile steepens. This is a natural result of the
increased acceleration of the neutrals. Second, the point where the
evolution becomes dynamical moves outward with time, and therefore the collapse
of the {\it envelope\/} is from inside-out {\it after\/} the collapse
of the core (see also Foster \& Chevalier, 1993\markcite{fc93}).
For our fiducial solution this collapse front moves at a
speed $\sim 0.05\,{\rm km\,s^{-1}}\,
(\rzero/0.5\,\pc)\,(n_{c,0}/10^3\,\ccc)^{1/2}$---smaller by $\sim 4$ than
a typical speed of sound in dark clouds.

\subsection{Mass Accretion Rates}

        We can use our results to evaluate the mass flow rates
in regions where the flow is quasi-static, and to estimate
the accretion rate onto the protostar after it forms.
In spherical symmetry the mass accretion rate at
any point $r$ in the cloud is given by
\begin{equation}
  \label{eq:mdot1}
  \dot{M}  =  4\pi r^2\,v_n\,\rho 
           \equiv  \dot{M}_{\rm AD}\, m\rhohat^{1/2},
\end{equation}
where we have used the non-dimensional expressions in
eqs. \refeqm{18}, \refeqm{19}, \refeqm{24a}, and \refeqm{29}.
The characteristic accretion rate due to ambipolar diffusion is
\begin{eqnarray}
  \label{eq:mdot0}
  \dot{M}_{\rm AD}\,&\equiv & \frac{\mcloud}{\tad} \nonumber \\
&=&  4.1\times10^{-7}\,\msolyr\,\left(\frac{\mcloud}{5\,\msol}\right)^{3/2}\, 
                             \left(\frac{R_0}{0.5\pc}
                              \right)^{-3/2}\, \epsilon^{-1/2}
\end{eqnarray}
where we have used equations \refeqm{23} and \refeqm{25}; recall
that $\epsilon$ is the initial ratio of mean to central density.
Figure 15 is a plot of $\dot{M}$ (in units of the fiducial mass
accretion rate $\dot{M}_{\rm AD}$) 
as a function of radius in our fiducial cloud for
both the pre- and post-collapse stages.

        Focusing first on the pre-collapse evolution, we
note that for our fiducial cloud ($\epsilon=0.1$) and reasonable values of the cloud's mass ($\mcloud\sim$ a few
\msol) and initial radius ($\rzero\la 1\pc$) the
value of $\dot{M}$ for most of the cloud is $\sim {\rm few}\,\times
10^{-7}\,\msolyr$. However, at any given time $t/\tad < 2$,  $\dot{M}$
has a maximum as a function of $r$; this maximal value increases
with time, and its location moves inward with
$\dot{M}(r=0)\rightarrow \infty$ as $\tau\rightarrow 2$.
 This feature is a direct
result of our assumed initial condition, which results in a constant
core mass as a function of time.

        As for the post-collapse accretion rates, 
$\dot{M}(r)$ is plotted in Figure 15 up to the
point where the quasi-static approximation breaks down ($\alpha\geq
0.1$). 
The results in Figure 15 show, again, that  to a good degree of
approximation the mass accretion
rate in the quasi-static envelope ($r/\rzero \ga 0.1$) is constant---with 
$\dot{M}\sim {\rm few}\,\times 10^{-7}\,\msolyr$. 

        How do these results depend on the free parameters, namely \mcloud,
\rzero, and $\epsilon$? Using equation~\refeqm{mdot0},
equation \refeqm{mdot1} can be rewritten as
\begin{equation}
  \label{eq:mdot2}
  \dot{M}  = 4.1\times10^{-7}\,\msolyr\,
                 \left(\frac{\mcloud}{5\,\msol}\right)^{3/2}\, 
                 \left(\frac{R_0}{0.5\pc}\right)^{-3/2}\, 
                  \epsilon^{-1/2} m\rhohat^{1/2}.
\end{equation}
We find that in the range $0.01\leq \epsilon \leq 0.2$ the combination
$\epsilon^{-1/2}\,m\rhohat^{1/2}\sim 0.6$--$1.6$ (depending on the
values of $r/\rzero$ and $t/\tad$ where the expression is evaluated);
this translates---holding \mcloud\ and \rzero\ constant---into a
change $\la 3$ in the mass accretion rates. Therefore, we conclude
that for  reasonable choices of the initial degree of concentration
($0.01\leq \epsilon \leq 0.2$), the cloud's mass ($\mcloud\sim$ a few
 solar masses), and initial cloud radius ($\rzero=0.5$--$1.0\pc$)
the mass accretion rate in the quasi-static envelope is $\sim {\rm
  few}\,\times 10^{-7}\,\msolyr$.

        Finally, we address the question of the mass accretion rate onto the
protostar, $\dot{M}_\ast$. 
By continuity, one would expect $\dot{M}_\ast$
 to be comparable to the mass accretion rate
 in the quasi-static envelope.  We can
answer this question more quantitatively with the help of our
expression for $r(t,M)$ (see eq.~\refeqp{p3}). Setting
 $\rhat=0$ in
equation~\refeqm{p3}, we can solve for the instantaneous value of $m$ at the
origin, i.e., the protostellar mass as a function of time\footnote{For
  $\tau \ge 2$---and as a result of ignoring thermal
  pressure---$\rhat=0$ is a singular point of
  equation~\refeqm{27}. However, we can still find $M_\ast(\tau)$ from
  equation~\refeqm{p3} by setting $\rhat=0$ because a real protostar will
  have a finite radius. For a typical protostellar radius
  $R_\ast\sim 5\,\rsol$ and an initial cloud radius $\rzero\sim 0.5\,\pc$, the
  lhs. of equation~\refeqm{p3} evaluated at the protostellar boundary  is
  of order $10^{-20}$---or zero, for all practical purposes.}; and by
computing its time derivative, obtain $\dot{M}_\ast(t)$.
Since we cannot treat the dynamical stage of evolution, this procedure
will become reasonably accurate only after about one free fall time
after the collapse.  However, although we cannot treat the rate
of accretion accurately during this stage, we can obtain a good
estimate for the integrated accretion--i.e.,
the total mass of the protostar that forms during the dynamical
phase. Figure 16 is a plot of $M_\ast$ and $\dot{M}_\ast$ (in units of,
respectively, \mcloud\ and $\dot{M}_{\rm AD}$) as a function of time after
the formation of the protostar for various values of $\epsilon$. 

 In our model, the formation of the initial protostar
corresponds to an infinite accretion rate. For $\tau-2
\ll 1$, we find that $M_\ast/M_0 -1 \propto (\tau-2)^{1/2}$, and therefore
$\dot{M}$---and the accretion luminosity---diverge as
$(\tau-2)^{-1/2}$ (though this is an integrable singularity, and the
total energy radiated by the shock vanishes as $(\tau-2)^{1/2}$ ).
However, this is an artifact due to
neglecting thermal stresses. Recall that in our model the protostellar
mass is a discontinous function of time near $\tau=2$, with $M_\ast =
 0$ as $\tau\longrightarrow 2^-$ and $M_\ast =
 m_c\,M_0$ as $\tau\longrightarrow 2^+$; the infinite accretion rate
 is a result of this discontinous behavior. When thermal pressure  is
 included, it provides a retarding force whose effect makes
  the protostellar mass a continous
  function of time at $\tau=2$. The calculations of Foster \&
  Chevalier (1993) illuminate this point. They followed the collapse
  of a non-singular cloud when only thermal pressure opposes gravity;
  therefore, their results are directly applicable to our phase of dynamical
  infall near $\tau=2$. They find that $\dot{M}$ is finite at the
  instant the protostar is formed, and it rises very sharply
  immediately thereafter. After this initial rise, which lasts for
  about a tenth of a free-fall time, the accretion rate declines in
  very much the same way as in our model. 

For our fiducial choice of parameters, a
 tenth of a free-fall time at the density of dynamical-infall onset
corresponds to $(\tau-2)\approx 2\times 10^{-3}$, or $t\al 10^4\,{\rm
 yr}$ after the formation of the initial protostar. Assuming a radius
 for the hydrostatic protostellar core $R_\ast=5\,R_\odot$, we find
an accretion luminosity 
\begin{equation}
  \label{eq:lacc}
  L_{\rm acc} = 230\,L_\odot\, m_c^2\, \epsilon^{-1/2}\,
  \left(\frac{\tau-2}{2\times 10^{-3}}\right)^{-1/2}\, 
  \left(\frac{M_0}{5\,M_\odot}\right)^{5/2}\, 
  \left(\frac{R_0}{0.5\,{\rm pc}}\right)^{-3/2}\, 
  \left(\frac{R_\ast}{5\,R_\odot}\right)^{-1}.
\end{equation}
For our fiducial model ($\epsilon=0.1$ and $m_c=0.26$, which
corresponds, for $M_0=5\,M_\odot$, to an initial protostar of
$1.3\,M_\odot$), we find that the accretion luminosity $\approx
10^4\,{\rm yr}$ after the formation of the initial protostar is
$49\,L_\odot$.  For an initial protostellar mass $M_\ast=0.5\,\msol$
(obtained with $\epsilon=0.025$, which gives $m_c=0.1$), we find, for
the same time after the initial collapse, an
accretion luminosity $\approx 15\,L_\odot$.
This last value is only a factor $\sim 2$ higher than those of typical
Class 0 sources (Gregersen et al., 1997\markcite{gal97}).

As shown in Figure 16, the mass
accretion rate remains very high for about a free fall time
($\sim 0.2\,\tad$) after the collapse, and, for the range in
$\epsilon$ explored, the protostellar mass increases by a factor $\sim
2$ during this phase. Thereafter,
$\dot{M}_\ast$ is comparable to the mass accretion rates in
the quasi-static envelope, as expected.
This strong dependence of
$\dot{M}_\ast$ shortly after the formation of the protostar has also
been found by Foster \& Chevalier (1993\markcite{fc93}) and Hunter
(1977\markcite{hun77}) for the case of non-magnetic collapse---a
result that is at strong variance with the predictions from the
singular isothermal sphere model (Shu 1977\markcite{shu77}).
The later stages of accretion are more in accord with the constant
accretion rate expected in the singular isothermal sphere model;
particularly---when one takes into account that
 $\dot{M}_{\rm AD} \propto \epsilon^{-1/2}$ 
(eq.~\refeqp{mdot0})---$\dot{\mstar}$ is essentially 
independent of $\epsilon$ during the
post-collapse, quasi-static phase. This similarity with the singular
isothermal sphere model merits further attention, and we address it in
the next section

\subsubsection{The Post-Collapse Mass Accretion Rate at Late Times \\ and
  the Singular Isothermal Sphere Model}

        The standard paradigm for the gravitational collapse of
an interstellar cloud is the inside-out collapse of a singular
isothermal sphere (Shu 1977).  The cloud is assumed to be
initially in an unstable equilibrium in which gravity is
balanced by the pressure of an isothermal gas
so that $\rho=c^2/2\pi Gr^2$, where $c$ is the isothermal
sound speed.  Collapse begins
at the origin, where the dynamical time is the smallest, and
an expansion wave moves outward at $c$.  At a time $t$ after
collapse begins, a mass $2c^3t/G$ has been swept up by the 
expansion wave; of this, a mass $0.975c^3t/G$ is in a condensed 
core at the origin.

In light of our results for the post-collapse, quasi-static accretion 
rate---namely, that at late times after
the collapse of the inner core of a non-singular stratified cloud the
mass accretion rate onto the protostar approaches a constant
value---we wish to compare our predicted accretion rates at late times
with that predicted by the singular isothermal sphere model.
To do this, we note that at $\tau=2$ the density distribution of the
envelope is given by $\rho\propto r^{-1/2}$ (see Figure 14), the same
density distribution as the singular isothermal cloud. At late times
($\tau\gg 1$),
the mass shells that still are in quasistatic equilibrium satisfy
$m\gg \mc$; 
in this limit, 
equation \refeqm{3.3}
for $m(r)$ reduces to
$m=\hat r$, and the initial density distribution 
(eq.~\refeqp{alphadef}) after core collapse 
becomes $\hat\rho(\tau=0)=(\mc/m)^2$.  
Rather than measuring time in units of the central
ambipolar diffusion time $t_{\rm AD}$, 
we use now the ambipolar diffusion time
at the surface, $t_{\rm AD}^\prime=t_{\rm AD}[\rho_{c0}/\rho(0,1)]^{1/2}
=t_{\rm AD}/\mc$.  The density distribution becomes
\begin{equation}
{\rho(\tau^\prime,m)\over \rho(0,1)}={4\over [m-\tau^\prime+(m^2-2m\tau^\prime)^{1/2}]^2}
        ~~~~~(m\gg \mc),
\label{eq:sing.rho}
\end{equation}
where $\tau^\prime\equiv t/t_{\rm AD}^\prime=\mc\tau$ and $\tau^\prime=0$ corresponds to
the formation of a singularity at the origin.
The radius of the mass shell $m$ is
\begin{equation}
\hat r^3=\frac12 m^3-\frac32 m^2\tau^\prime+\frac34 m\tau^{\prime\,2}+\frac12
        (m^2-2m\tau^\prime)^{3/2}~~~~~(m\gg \mc).
\label{eq:sing.r}
\end{equation}
This equation indicates that the shell reaches the origin at
a time $\tau^\prime_{\rm cr}=\frac49 m$.  By this time, our approximations for
the dynamics have broken down, but this nonetheless defines 
a characteristic time scale for the evolution.  At this time,
equation (\ref{eq:sing.rho}) 
shows that the density has increased by a factor of
about 5.

        If we approximate
the acceleration of the gas by $v_n^2/r$, the parameter $\alpha$
(eq.~\refeqp{alphadef}) becomes
\begin{equation}
\alpha\simeq {4\pi G\over k_{ni}^2}\left[\frac{r(0,m)}{r(\tau^\prime,m)}\right]^3
        \left[\frac{\rho(0,m)}{\rho(\tau^\prime,m)}\right].
\end{equation}
As discussed in \S 5.2.1, $\alpha$ is initially very small; it
will begin to approach unity (so that our quasi-static approximation
breaks down) only when $\hat r^3$ becomes significantly smaller than
its initial value.  Equation (\ref{eq:sing.r}) shows that this
will occur only when $\tau^\prime$ is quite close to $\tau^\prime_{\rm cr}$.
It is therefore self-consistent to use to use $\tau^\prime_{\rm cr}$
as an estimate for the evolution time for the shell.

        The time scale for evolution $\tau^\prime_{\rm cr}$ enables
us to define the characteristic velocity with which ambipolar
diffusion
advances into the cloud,
\begin{equation}
v_{\rm AD}\equiv \frac{r(0,m)}{\tau^\prime_{\rm cr} t_{\rm AD}^\prime}
        =\frac94\frac{R_0}{t_{\rm AD}^\prime},
\end{equation}
which is constant.  How does this compare with the 
expansion wave associated with the collapse of
an unmagnetized, singular isothermal sphere, which moves outward
at the sound speed $c$?   Defining an
effective sound speed through the relation governing such a 
sphere, $c^2=2\pi r^2G\rho$, and using equation (40), we find
\begin{equation}
v_{\rm AD}=\frac{3.2c}{\nuff }.
\label{eq:vad}
\end{equation}
Thus, for typical conditions in a shielded core in which
the ionization is dominated by cosmic rays, so that $\nuff \sim 10$, 
the ambipolar diffusion advances into the envelope at about
1/3 the speed of the expansion wave.
The result is not changed significantly if we instead
compare our result with that for the collapse of a
thin disk with a frozen-in magnetic field: Li \&
Shu (1997) have shown that in this case the expansion
wave advances at $c\surd 2$,
about the same velocity as in the unmagnetized
case.

        The accretion rate is given by equation (101).
Evaluating this rate at the time $\tau_{\rm cr}^\prime$, we find
\begin{equation}
\dot M=\frac94\left(\frac{M_0}{t_{\rm AD}^\prime}\right)
        =5.4\times 10^{-7}\left(\frac{M_0}{5\;M_\odot}\right)^{3/2}
        \left(\frac{R_0}{0.5\;{\rm pc}}\right)^{-3/2}~~~~~
        M_\odot\;{\rm yr}^{-1},
\label{eq:mdotad}
\end{equation}
in good agreement with the estimate for the non-singular case
in \S 5.7.  If we express the accretion rate in terms of
the sound speed required to support the sphere, as we did
in equation (\ref{eq:vad}), we find
\begin{equation}
\dot M=\frac{6.38}{\nuff }\left(\frac{c^3}{G}\right).
\label{eq:mdotadsis}
\end{equation}
Recall that for the non-magnetized singular isothermal sphere,
the accretion rate is $\dot M=0.975 c^3/G$ (Shu 1977).
Our results indicate that when ambipolar diffusion is included,
the accretion rate is reduced only slightly:
for $\nuff \sim 10$, we obtain an accretion rate
that is about 2/3 that for the equivalent non-magnetized case.

        This result is quite remarkable: it indicates that
the presence of a strong magnetic field does not significantly
reduce the accretion rate onto a protostar 
at late times
in a cloud in which
the ionization is due to cosmic rays (photoionization by FUV or
X-ray photons increases $\nuff $ and reduces the accretion
rate proportionately).  To see why this is so, let us first
estimate the accretion rate for the singular isothermal sphere.
Since about half the swept up mass has collapsed to the origin
at any time, the accretion rate is $\dot M_{\rm SIS}\sim M/2\tff$,
where $\tff$ is the free fall time based on the mean
density $\bar\rho=3\rho_0$.  In terms of 
$\tff^\prime$, the free fall time at $\rho_0$, we then have
\begin{equation}
\dot M_{\rm SIS}\sim\frac{\surd 3}{2}\left(\frac{M}{\tff^\prime}\right)=
        \frac{4}{\pi}\left(\frac{c^3}{G}\right).
\end{equation}
This is somewhat higher than the exact answer, presumably because
we neglected the retarding effects of the gas pressure.  

        Next, consider the magnetized
case.  For a uniform cloud, the average accretion
rate is $\dot M\sim M/2\tad$ based on the results in \S 4.  
Since the mass
of a singular sphere is 3 times that of a uniform sphere with
the same density at the surface, the ambipolar diffusion velocity,
and hence the collapse rate, will be three times greater:
$\dot M_{\rm SIS,AD}\sim \frac32(M/\tad^\prime)$, which is within a factor 1.5 of
the exact answer in equation (\ref{eq:mdotad}).  With these
estimates, we find
\begin{equation}
\frac{\dot M_{\rm SIS,AD}}{\dot M_{\rm SIS}}\sim
        \frac{3M/2\tad^\prime}{\surd 3 M/2 \tff^\prime}=\surd 3
        \left(\frac{\tff^\prime}{\tad^\prime}\right)=\frac{\surd 3}{\left(\frac{8}{3\pi^2}\right)^{1/2}\,\nuff }.
\end{equation}
With allowance for the fact that our estimate for this ratio
is off by a factor 2 (the estimate for the magnetized case was
somewhat too low, and that for the unmagnetized case was too high),
this agrees with the result in equation (\ref{eq:mdotadsis}).
Several factors have raised the magnetized
accretion rate close to the unmagnetized rate: in the magnetized case,
the entire cloud is contracting at $t=2\,\tad$, whereas in the
unmagnetized case the collapse begins only after the expansion wave
reaches the point in question; thermal pressure retards the unmagnetized case
somewhat (we have assumed that magnetic pressure dominates thermal
pressure in the magnetized case); and finally, in going from an
estimate
based on a uniform sphere to a singular sphere, the rate for
the magnetized case increases by $\bar\rho/\rho_0=3$, whereas
that for the unmagnetized case increases only by the square root
of this ratio.

        We conclude that, contrary to one's expectation
(Mouschovias 1991),
the accretion rate onto a protostar at late times after its formation 
is not significantly inhibited
by the presence of a strong magnetic field, so long as the ionization
is due to Galactic cosmic rays. 
Note that this result does not
validate the singular isothermal model {\it in toto}. It only means
that at late times, {\it after\/} the formation of the protostar, the
mass accretion rate onto the central object approaches the limiting
case of a singular isothermal cloud. We must stress, again, that our
results for core collapse and immediately thereafter are at strong
variance with the predictions of the singular isothermal sphere model.

In our idealized model, the  accretion will continue until the 
mass of the envelope is exhausted.  In reality, several other 
factors could intervene:
the outer envelope is likely to be photoionized (McKee 1989), with
a much longer ambipolar diffusion time,
and the outflow
from the protostar could disperse the envelope (Shu et al 1987).

\section{Conclusions}

In this work we have investigated the evolution of a cold, magnetized
molecular cloud by means of an idealized, semi-analytic model.
We neglect thermal pressure, assume quasi-static evolution,
and impose spherical symmetry. Our
principal conclusions are the following:
\begin{enumerate}
\item In the absence of thermal stresses, an initially homogeneous
  collapses homologously, without developing any structure.
  The collapse reaches a singularity at a time $2\,\tad$,
  where \tad\ is the initial  ambipolar diffusion time.
\item The central core of a centrally condensed 
  cloud also collapses after a time $2\,\tad$ (where $\tad$ is
  measured at the center of the cloud),
  leaving behind a quasi-static envelope.  
\item In a centrally condensed cloud 
  the collapse of the core proceeds from {\it outside-in}, while
  the quasi-static envelope collapses from {\it inside-out\/} {\rm
  after} the collapse of the central region.
\item The pre-collapse evolution of the core 
  is similar to the evolution
  of a homogeneous cloud. In particular, the quasi-static 
  evolution of the central density is a universal function 
  of time, $\rho_c \propto (2-t/\tad)^{-2}$, independent of any parameters
  that enter the calculation.
\item The core evolves quasi-statically up to the point that it
  becomes magnetically supercritical.  Thereafter, the ions are no
  longer approximately static, and our approximations break down.
  As a result, we cannot treat the subsequent dynamical collapse.
  Nonetheless, even after the formation of the central protostar,
  we are able to follow the quasi-static evolution of the envelope
  so long as it remains magnetically subcritical. 
\item Most of the initial magnetic flux 
  in the core is lost via ambipolar diffusion; after collapse,
  we estimate that
  the remaining trapped flux is about half the critical value.
\item The mass accretion rate in the quasi-static envelope before and
  after core   collapse  is   remarkably constant and of order
  \begin{displaymath}
      \dot{M}  \sim {\rm few}\,\times 10^{-7}\,\msolyr\,
                  \left(\frac{\mcloud}{5\,\msol}\right)^{3/2}\, 
                  \left(\frac{\rzero}{1.0\pc}\right)^{-3/2}\, 
                  \left(\frac{\epsilon}{0.1}\right)^{-1/2}\,,
  \end{displaymath}
  where \mcloud\ and \rzero\ are, respectively, the cloud's mass and
  inital radius and $\epsilon$ is the inital ratio of mean to central
  density, with $0.01\la \epsilon \la 0.2$.
\item  Much of the mass of the star is built up during the
  dynamical phase of accretion, which lasts about a free-fall
  time.  Although we cannot treat this dynamical phase,
  our model does enable us to determine the total mass accreted
  and to follow the subsequent quasi-static accretion.
  Just as in the non-magnetic case (Foster \& Chevalier 1993), 
  the accretion rate during this stage is much larger than that
  of a singular isothermal sphere.  
\item The late-time mass accretion rate onto the protostar  is
  quasi-static.  Notwithstanding the inhibiting effects of the magnetic
  field, the accretion rate becomes comparable to that of
  a singular isothermal sphere.
\item The excellent agreement between our results for the pre-collapse
  evolution and the detailed
  numerical work of Fiedler \& Mouschovias \markcite{fm92}(1992)
  indicates that our simple
  approach captures the basic physics relevant to the evolution of a
  cloud undergoing ambipolar diffusion. 
  Our results are especially relevant to the late stages of cloud
  evolution, when thermal stresses are relatively small.
  Since thermal pressure does play an important role at earlier times,
  particularly in creating the initial density distribution, our
  intention is to extend the present calculation to one that 
  incorporates thermal pressure in a subsequent paper.
\end{enumerate}

\acknowledgements

The research of PNS was supported by a NASA grant to the Center for
Star Formation Studies and by NSF grant AST9314847 to the Laboratory
for Millimeter-Wave Astronomy at the University of Maryland; the
research of CFM is supported by the
National Science Fundation (AST95--30480) and by a NASA grant to the
Center for Star Formation Studies; and SWS is supported by the NASA
Astrophysics Theory Program, grant NAGW--3107.

\clearpage
%
%
%

\figcaption{
Evolution of the Lagrangian radius $r$ (in units of the initial cloud
radius $R_0$) as a
function of time for a
homogeneous cloud with $\delta = 4\times 10^{-3}$.
The five curves show  the evolution of $r$ for five different
values of the
dimensionless Lagrangian coordinate $m$:
 $m=10^{-2}$, $3.16\times 10^{-2}$, $10^{-1}$, $3.16\times 10^{-1}$,
 and $1$. All the
masses are in units of the cloud's mass \mcloud, and time is in units of the
characteristic ambipolar diffusion time, \tad.
}

%
%
%
%
\figcaption{ Evoution of the magnetic critical ratio $\Mphi/M$ for a
  homogeneous cloud with  $\delta = 4\times 10^{-3}$.
 {\it (a)\/}: $\Mphi/M$ as a
  function of time (in units of the ambipolar diffusion time \tad) for
  three different values of the dimensionless Lagrangian coordinate
  $m$: $m=10^{-2}$, $m=10^{-1}$, and
  $m=1.0$. {\it (b)\/}: $\Mphi/M$ as  a function of the dimensionless
  coordinate $M/\mcloud$
at three different times (in units of the
  characteristic ambipolar diffusion time): $\tau=0.0$,
  $1.62$, and $1.86$.
}
%
%
%

\figcaption{ Ratio of the ion's velocity $v_i$ to the neutral's
  velocity $v_n$ for a homogeneous cloud with 
$\delta = 4\times 10^{-3}$. Shown is $v_i/v_n$ as a function of
  the dimensionless Lagrangian coordinate $M/\mcloud$  
  at four different times (in units
  of the   characteristic ambipolar diffusion time): $\tau=0.0$, $1.00$,
  $1.62$, and $1.86$.
}

%
%
%

\figcaption{
The mass of the initial central core, \mcore\ (in units of \mcloud), as a
function of the initial 
degree of concentration, $\epsilon=\left<\rho\right>/\rhoc$ for a stratified
cloud with $p=2$.
 Only values
$m_c=\mcore/\mcloud \le 1$ are 
physically relevant. Clouds with $\epsilon\ge 3/7$ 
have $\mc \ge 1$, and therefore collapse as a
homogeneous cloud.
}

%
%
%
\figcaption{Lagrangian radius $r$, neutral velocity
  $v_n$, density $\rho$, and magnetic field $B$ for our fiducial
  stratified cloud with an initial degree of concentration
  $\epsilon=0.1$ and magnetic parameter $\delta=4\times10^{-3}$ at five
  different times (in units of the characteristic ambipolar diffusion
  time \tad): $\tau=0.0$, $0.88$, $1.37$,
  $1.64$, and $1.80$ (between consecutive values of
  $\tau$ the central density increases by a factor of $(10)^{1/2}$).
{\it (a)\/}: Lagrangian radius $r$ (in units of the cloud's initial
  radius $R_0$--{\it top} panel) and neutral velocity $v_n$ (in units
  of the fiducial velocity $v_{n,0}$--{\it bottom} panel) as a function
  of the dimensionless Lagrangian coordinate $M/\mcloud$. Time increases from
  {\it top} to {\it bottom} for the $r$-curves and from {\it bottom}
  to {\it top} for $v_n$. The {\it dashed} line indicates the location
  of the central core. {\it (b)\/}: Density in units of the initial
  central density \rhoc\ ({\it top} panel) and
  magnetic field (plotted as
  $B/B_0\,-1$, where $B_0$ is the strength of the magnetic field at
  the edge of the cloud; {\it bottom}  panel)  as  functions of the
  dimensionless Lagrangian coordinate $M/\mcloud$.  Time increases from {\it
  bottom} to {\it top}. The {\it dashed} line indicates the location
  of the central core.
}

%
%
%
%

\figcaption{
The acceleration of the neutrals in units of the local gravitational
acceleration. Shown is $\alpha$, for our fiducial cloud, as a function
of the dimensionless Lagrangian coordinate $M/\mcloud$ and 
dimensionless reversed time (measured from the formation of a
singularity at the origin) $2-t/\tad$.
}

%
%
%
\figcaption{
Time evolution of the magnetic flux $\Phi$ in our fiducial cloud
 (in units of the fiducial
magnetic flux $\pi\bzero\,R_0^2$) as a function of the dimensionless Lagrangian
coordinate $M/\mcloud$ ({\it left} panel) and the {\it Eulerian}
radial coordinate $R$ (in units of the cloud's initial radius \rzero;
{\it right} panel). Each curve corresponds to a different
dimensionless time $\tau$ ( $\tau=0.0$, $0.88$, 
$1.37$,  $1.64$, and $1.80$), which increases
from {\it top} to {\it bottom} in the {\it left} panel and from  
{\it bottom\/} to {\it top\/} in the {\it right\/} panel. The {\it
  dashed\/} line and the arrows indicate the location  of the central
core; the arrows are labeled by the value of $t/\tad$.
}
%
%
%
%
\figcaption{
Ratio of the ion to neutral velocity, $v_i/v_n$ for our fiducial cloud
($\epsilon=0.1$ and $\delta=4\times 10^{-3}$), as a function of
the dimensionless Lagrangian coordinate $M/\mcloud$ for five different
times (in units of the characteristic ambipolar diffusion time \tad;
from {\it bottom} to {\it top}): $\tau=0.0$, $0.88$, 
$1.37$,  $1.64$, and $1.80$.  
The {\it dashed} line indicates the location  of the central core
}
%
%
%
%
\figcaption{ Evolution of the critical ratio $\Mphi/M$ for our fiducial
 cloud ($\epsilon=0.1$ and $\delta = 4\times
 10^{-3}$). $\Mphi/M$ is plotted as a function of the dimensionless
 Lagrangian coordinate $M/\mcloud$
for five different
times (in units of the characteristic ambipolar diffusion time \tad;
from {\it top} to {\it bottom}): $\tau=0.0$, $0.88$, 
$1.37$,  $1.64$, and $1.80$.  
}
%
%
%
%
\figcaption{ 
Ratio of thermal to magnetic stresses, $\beta^\prime$, as a function of
the dimensionless Lagrangian coordinate $M/\mcloud$ and 
dimensionless reversed time (measured from the formation of a
singularity at the origin) $2-t/\tad$.
 Shown are the results for our fiducial cloud, $\epsilon=0.1$ and
 $\delta=4\times 10^{-3}$, with 
initial central number density $n_{c,0}=2200\,{\rm cm^{-3}}$, 
magnetic field strength $\bzero=30\,\mu\gauss$ at the edge of the
cloud, and temperature $T=10\kelvin$.
}

%
%
%
%
\figcaption{
Comparison with the numerical results of Fiedler \& Mouschovias (1992) for
the time
evolution of the central number density  $n_c$ ({\it top\/} panel) 
and central magnetic field $B_c$ ({\it bottom\/} panel)
in a cloud with mass $\mcloud=45.5\,\msol$,
initial radius $\rzero=0.75\,{\rm pc}$, and external magnetic field
$\bzero=30\,\mu{\rm G}$. Shown are the results of Fiedler \&
Mouschovias for their Model~1 
({\it filled\/} circles and {\it dashed\/} line)
 after the initial relaxation and up
to the onset of dynamical contraction and those from our semi-analytic
model ({\it solid} line).
}

%
%
%
%
\figcaption{
Comparison with the numerical results of Fiedler \& Mouschovias (1992) for
the evolution of the density profile. Shown
is the  number density as a function of dimensionless radius
$r/\rzero$ for three different times during the quasi-static phase.
 The {\it filled} circles are the
numerical results for the midplane of Fiedler \& Mouschovias's Model 1
({\it light grey:\/} $t=10.2\, {\rm Myr}$, {\it dark grey:\/} $t=15.1\,
{\rm Myr}$, and {\it black:\/} $t=16.0\, {\rm Myr}$);
the {\it solid} lines are the results from our semi-analytic
model. For easier visualization, the curves have been shifted
horizontally by a constant $C$: $C=\log 0.5$ for $t=10.0\, {\rm Myr}$;
$C=\log 0.75$ for $t=15.1\, {\rm Myr}$; and $C=0.0$ for $t=16.0\,{\rm Myr}$.
}

%
%
%
\figcaption{Core's magnetic flux $\Phi_{\rm core}$ (in units of the
  cloud's initial magnetic flux $\Phi_0$) as a function of
  dimensionless reversed time (measured from the formation of a
  singularity at the origin) $2-t/\tad$. Shown is $\Phi_{\rm core}/\Phi_0$
  for $\delta=10^{-3}$ and four different values of $\epsilon$:
  $\epsilon=10^{-4}$, $10^{-3}$, $10^{-2}$, and $10^{-1}$.
}
%
%
%
\figcaption{Post-collapse evolution of the density profile ({\it top\/} panel)
  and neutral velocity profile ({\it bottom\/} panel)  for our
  fiducial cloud ($\epsilon=0.1$). Shown are the
  density in units of the initial central density and the neutral
  velocity in units of the fiducial velocity $v_{n,0}$ as a function of
  radius in the cloud (in units of the initial cloud radius \rzero)
  for six different times: $\tau=2.0$, $2.1$, $2.2$, $2.3$, $2.4$,
  and $2.5$. Time increases from the {\it topmost\/} curve {\it down\/}. The
  inner boundary of each curve corresponds to the point where the
  acceleration of the neutrals is $10$\% of the local gravitational
  acceleration. 
}
%
%
%
\figcaption{Mass accretion rates as a function of radius for our fiducial cloud
  ($\epsilon=0.1$). Shown is $\dot{M}$ (in units of the fiducial mass
  accretion rate $\dot{M}_{\rm AD}$) as a function of radius in the cloud (in
  units of \rzero) before the collapse of the inner core ({\it
  dotted\/} lines) and afterwards ({\it solid\/} line and {\it
  shaded\/} region). For the pre-collapse accretion rates each curve
  corresponds to a different time (in units of the characteristic
  ambipolar diffusion time \tad; from {\it bottom\/} to {\it top\/}):
  $\tau=0.0$, $0.88$, $1.37$, $1.64$, and $1.80$. The {\it solid\/}
  line is the accretion rate at the time when the core collapses
  ($\tau=2.0$) and the {\it shaded\/} region corresponds to the
  post-collapse mass accretion rates for $\tau\le 2.5$; the
  individual curves in this regime are too close to each other to show
  clearly on the scale of this plot. The post-collapse curves
  ($\tau\ge2.0$) are truncated to the left
 at the point where the acceleration of
  the neutrals is $10$\% of the local gravitational acceleration.
}
%
%
%
%
\figcaption{Protostellar mass \mstar\ and mass accretion rate onto the
  protostar, $\dot{M}_\ast$, as
  functions of dimensionless time (measured from the formation of a
singularity at the origin) $t/\tad-2$. Shown are
  $\mstar$ (in units of the cloud's mass \mcloud; {\it top\/} panel) and
  $\dot{M}_\ast$ (in units of the fiducial mass accretion rate
  $\dot{M}_{\rm AD}$; 
  {\it bottom\/} panel) for three
  different values of $\epsilon$: $\epsilon=0.05$, $0.1$, and $0.2$.
}
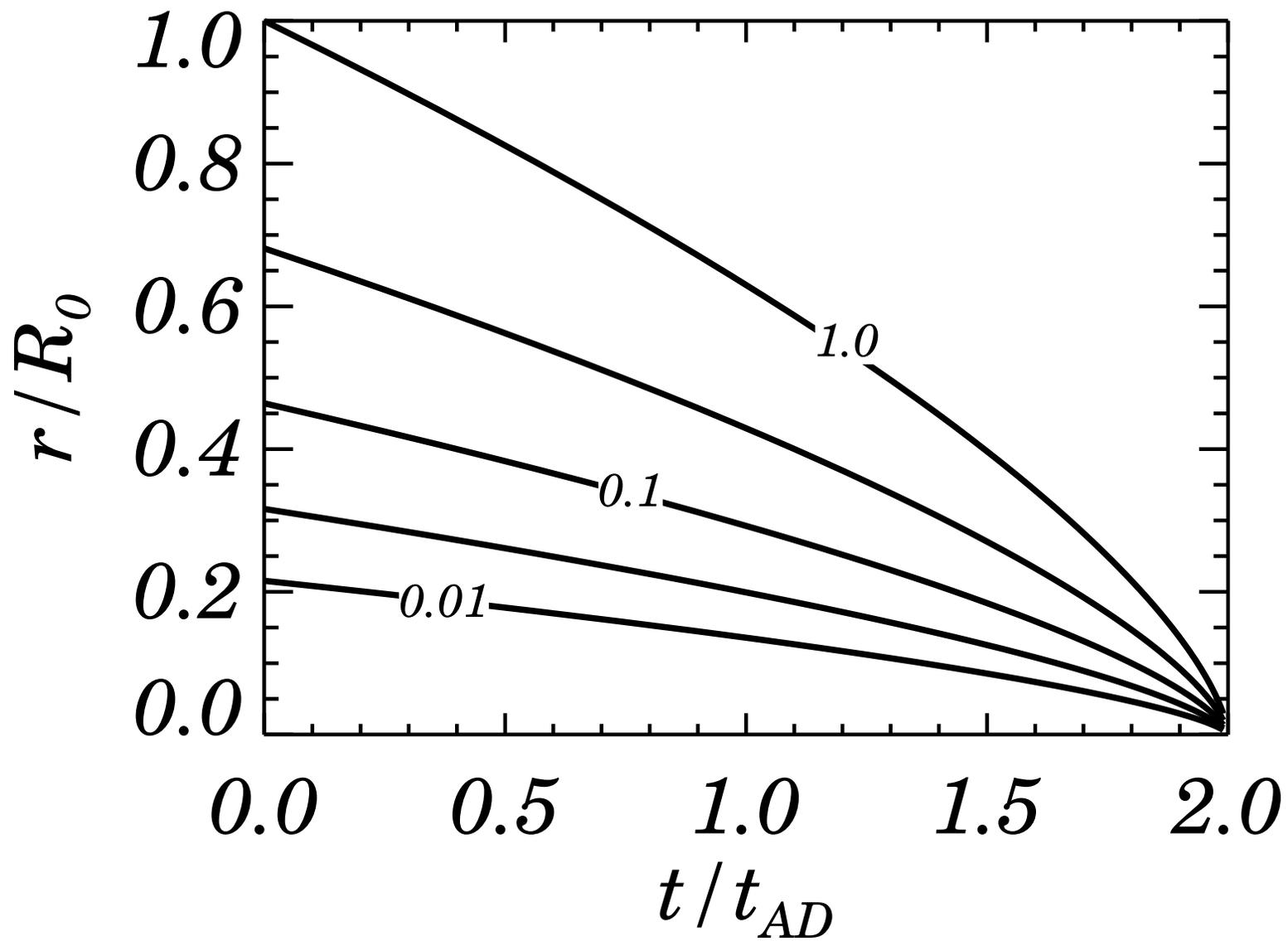
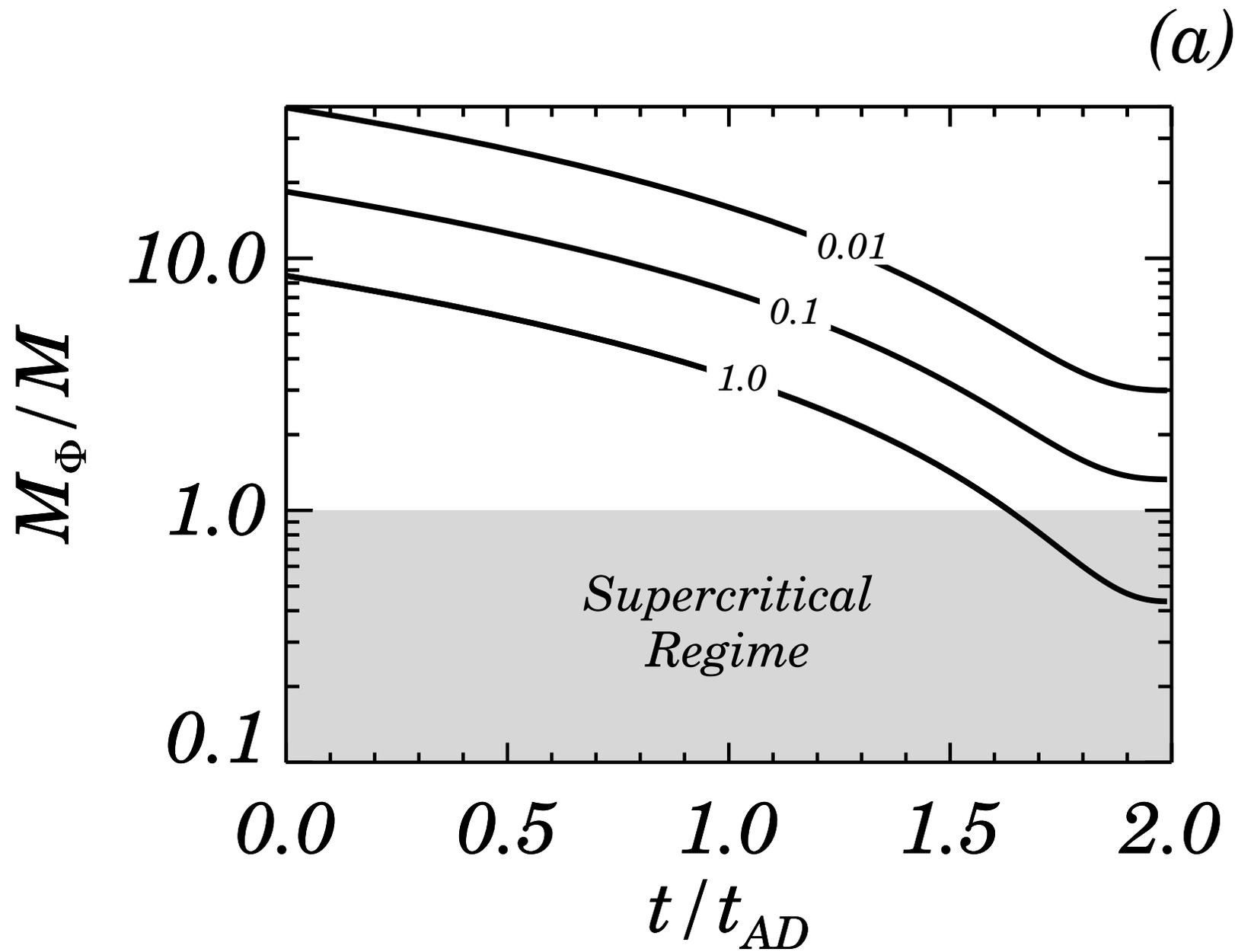
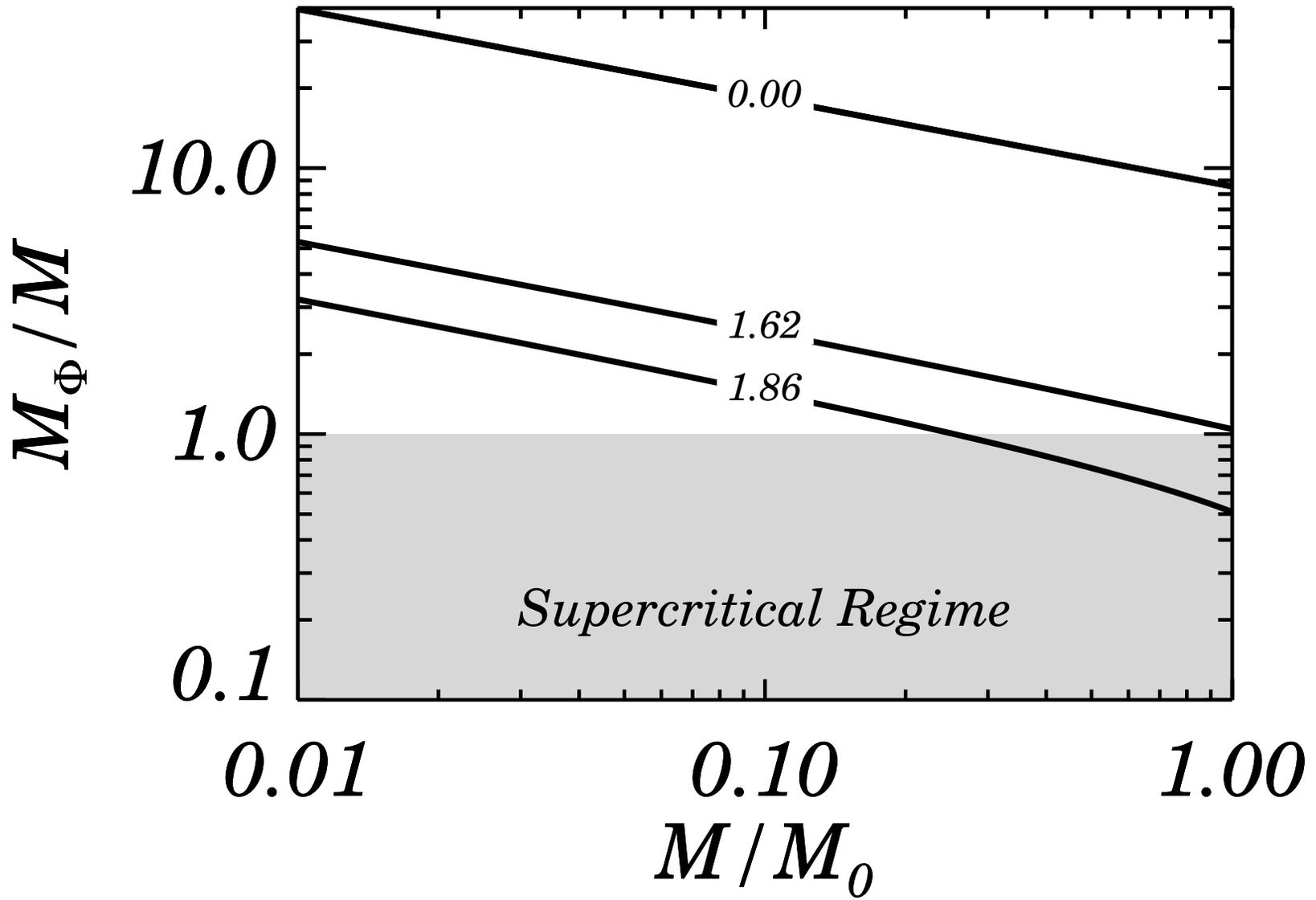
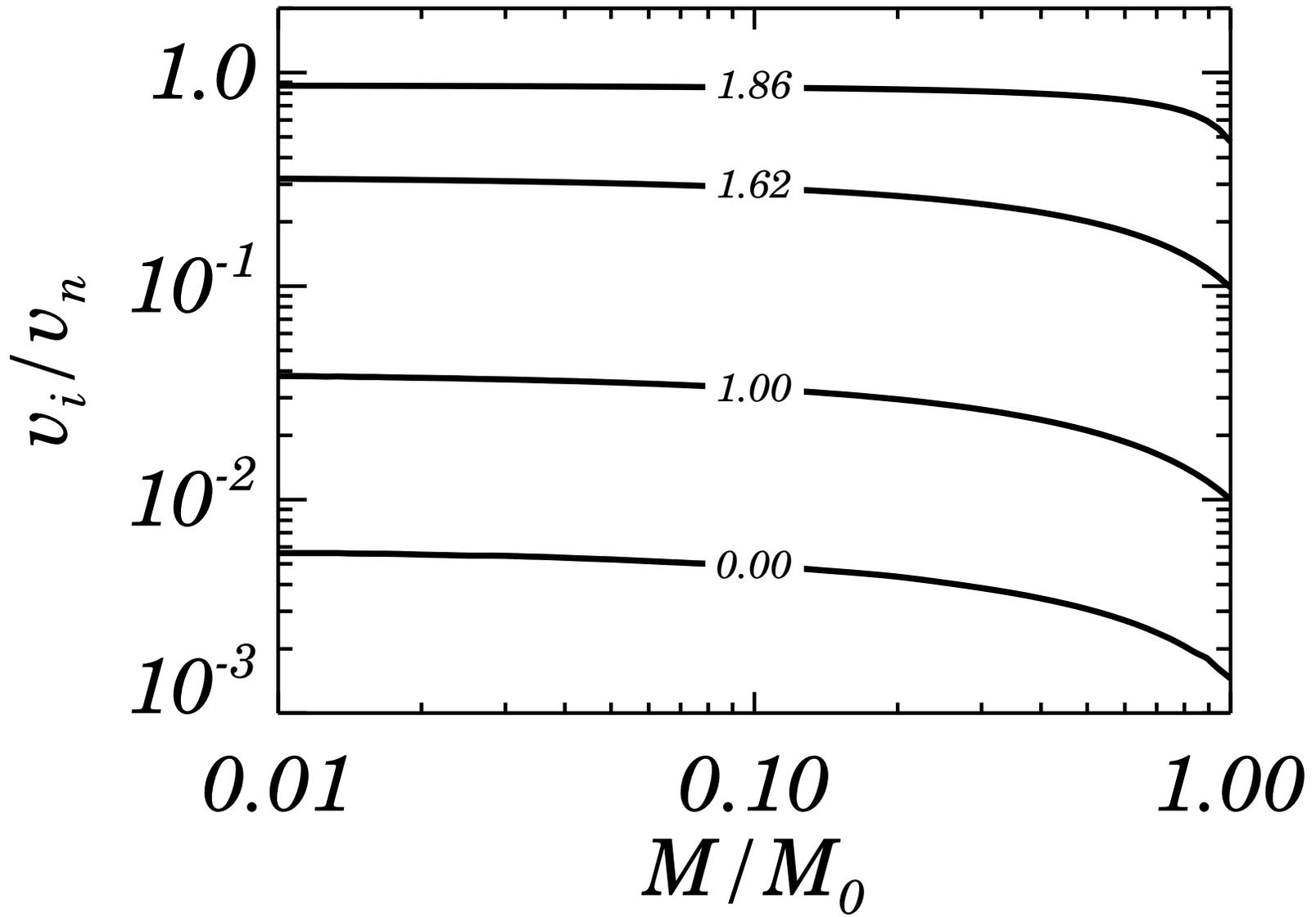
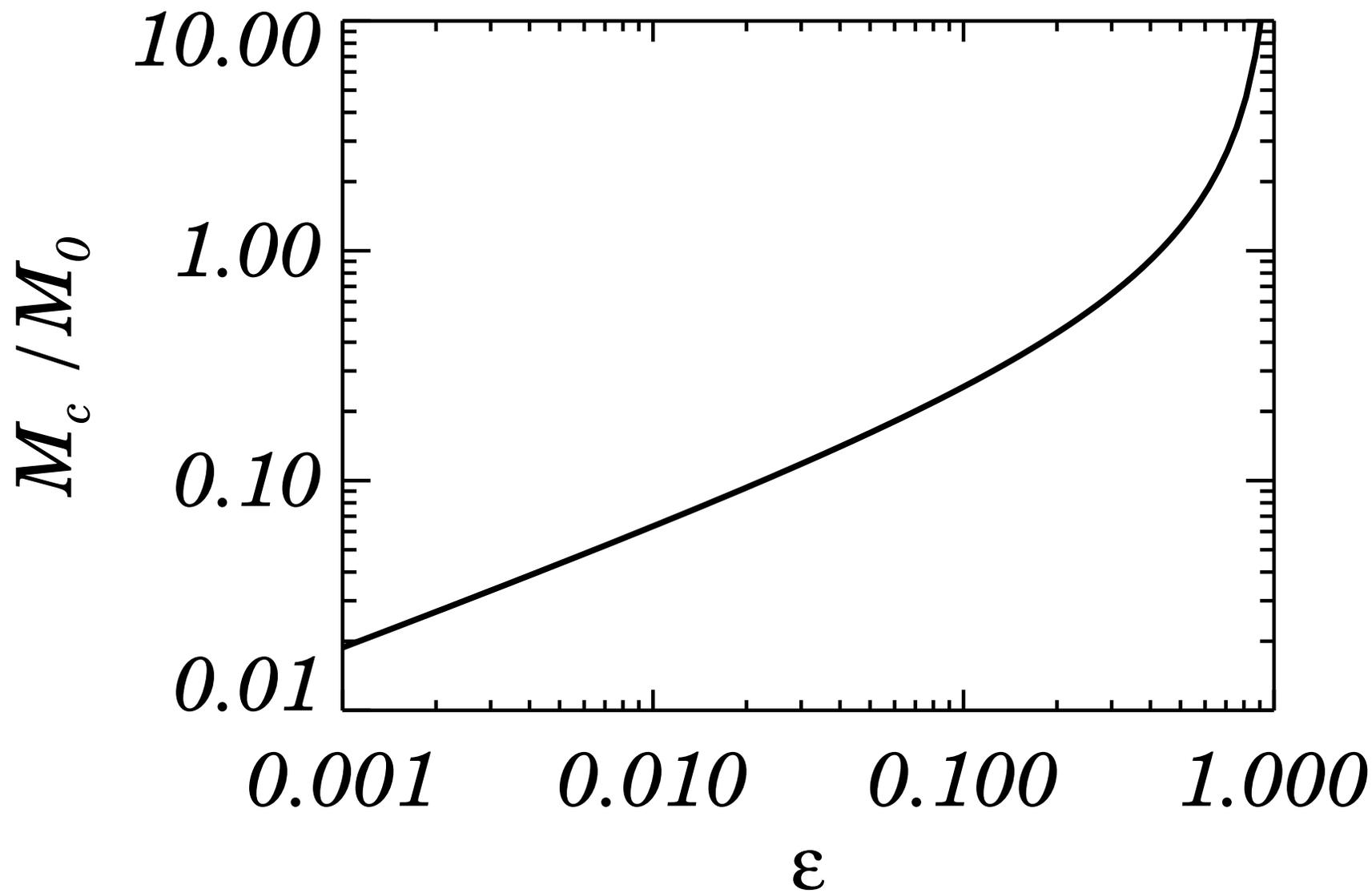
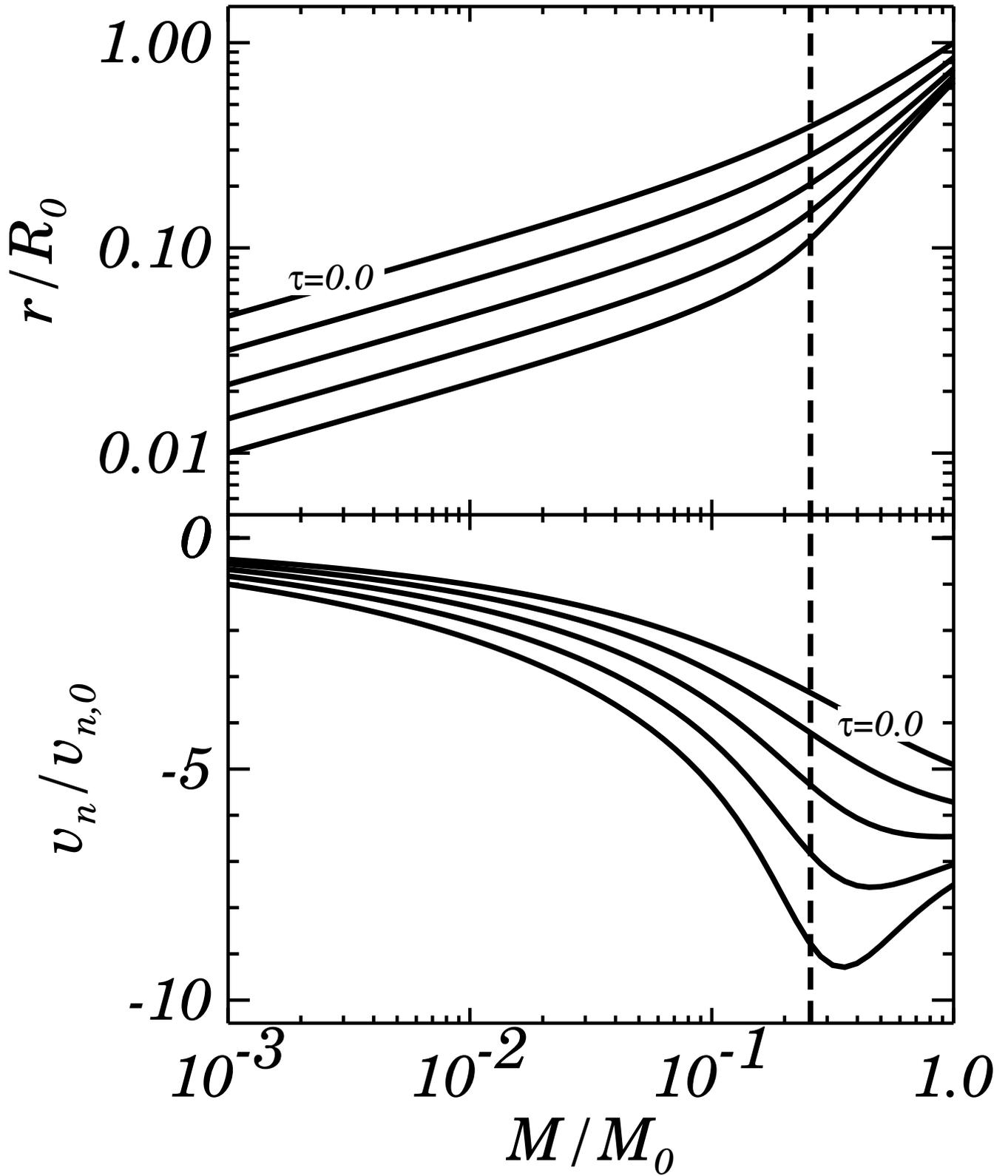
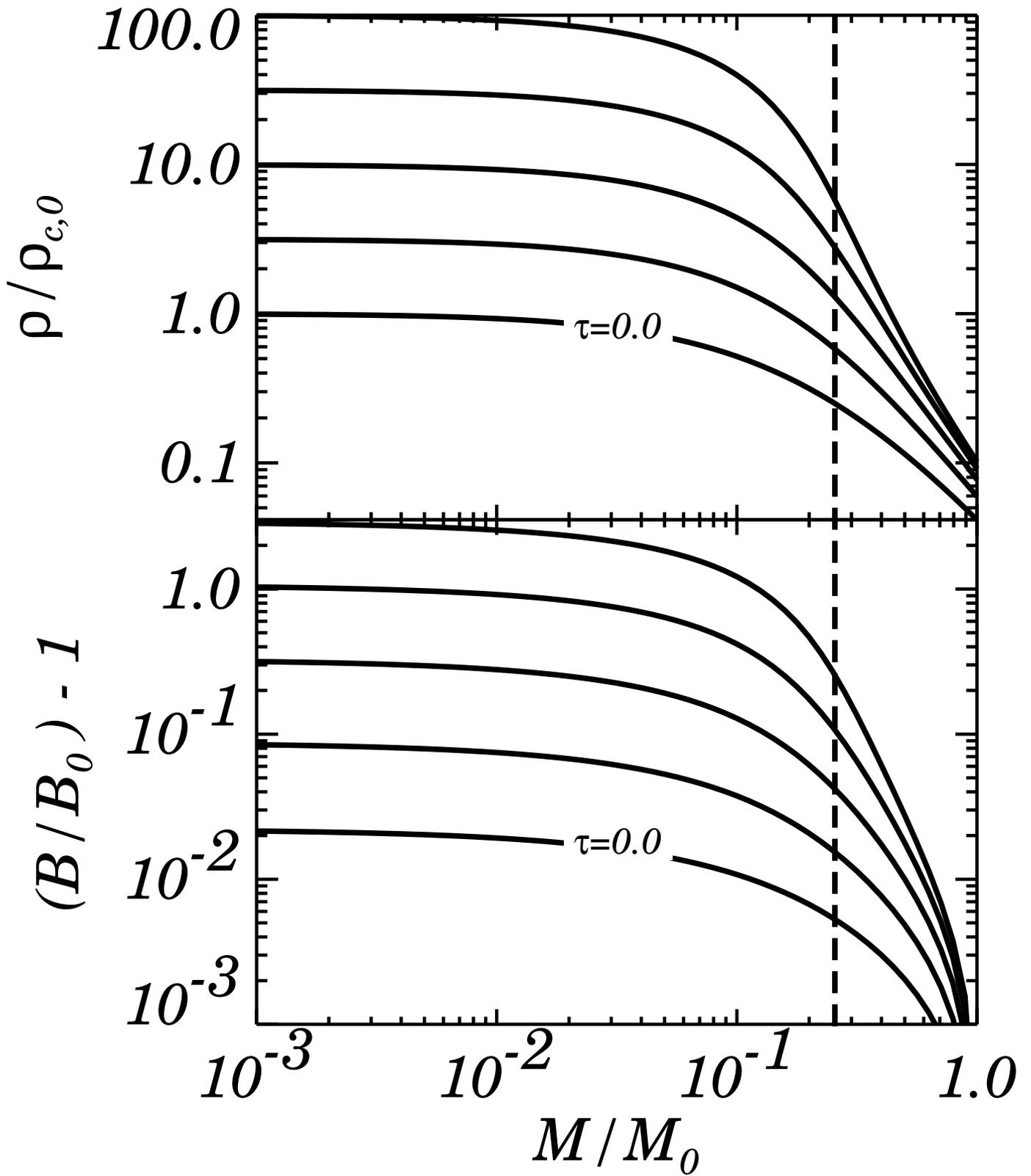
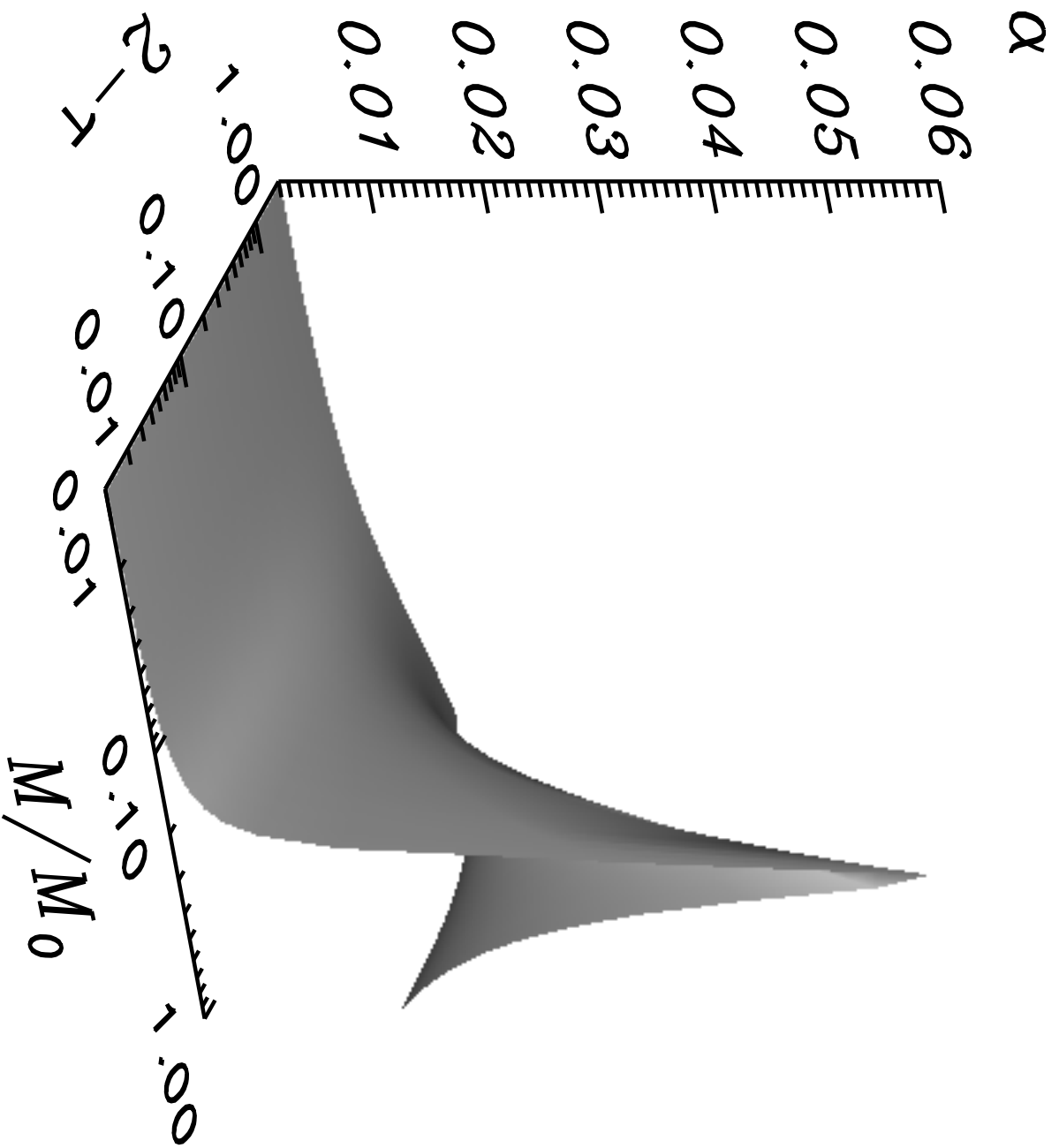
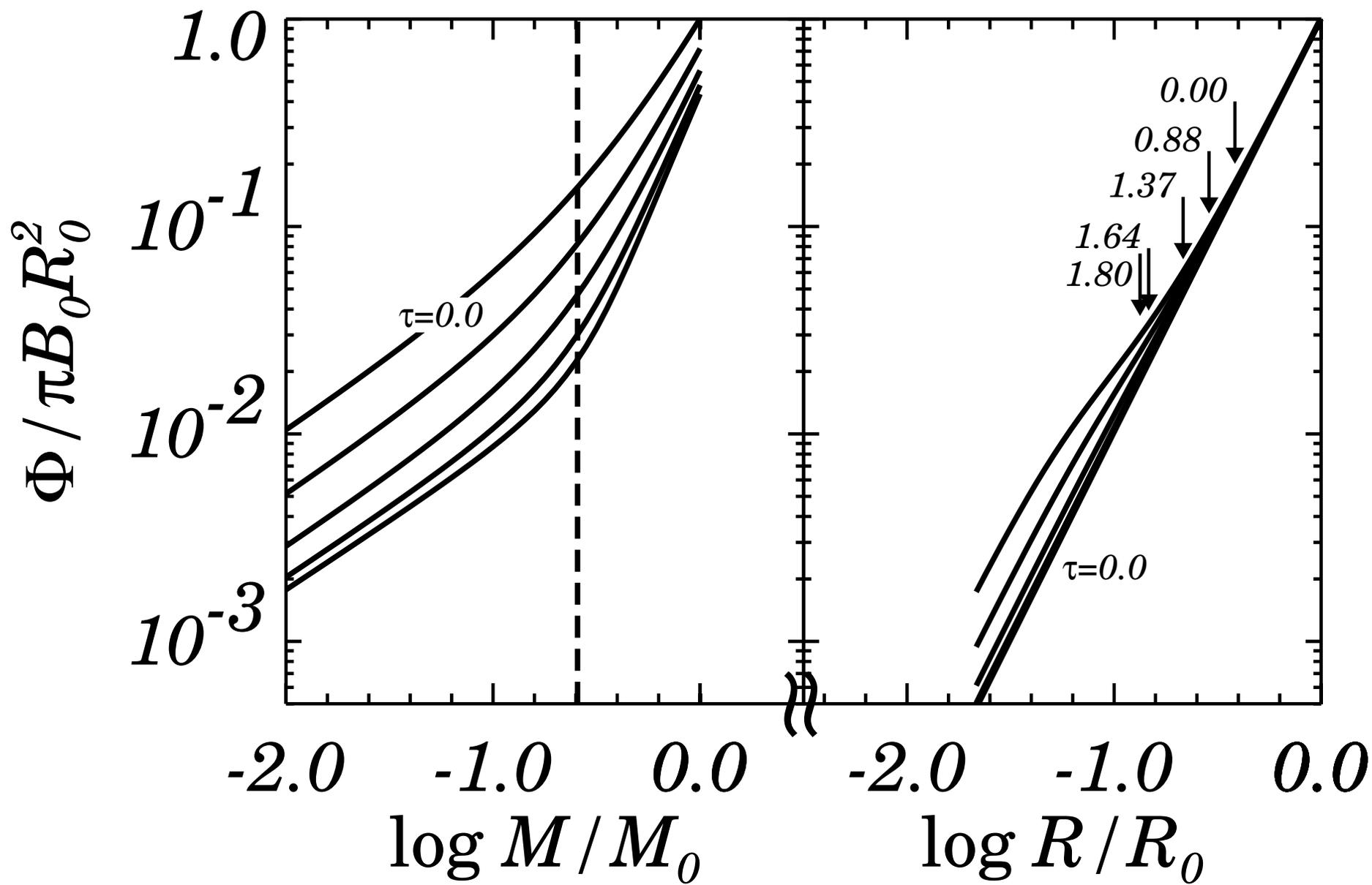
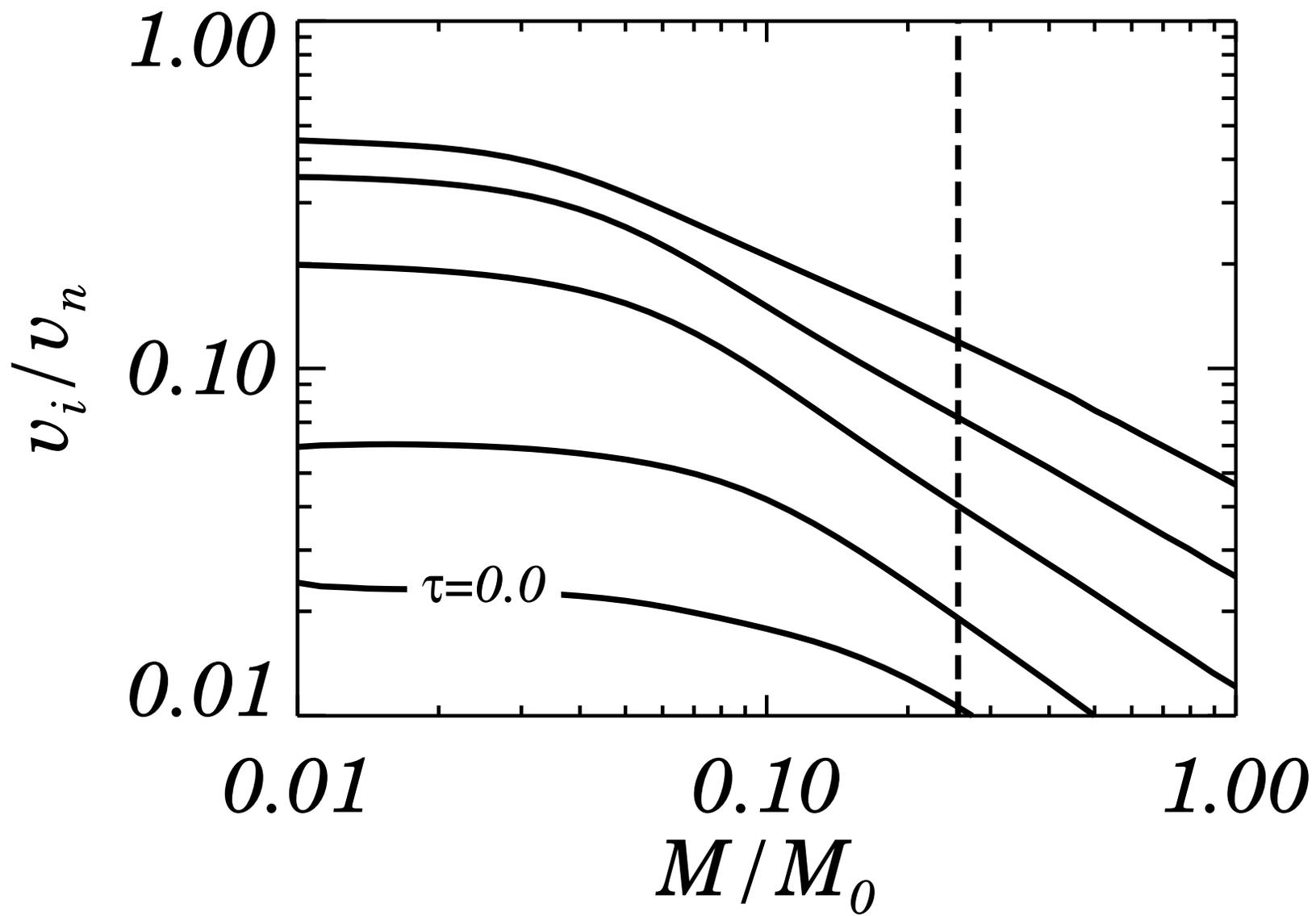
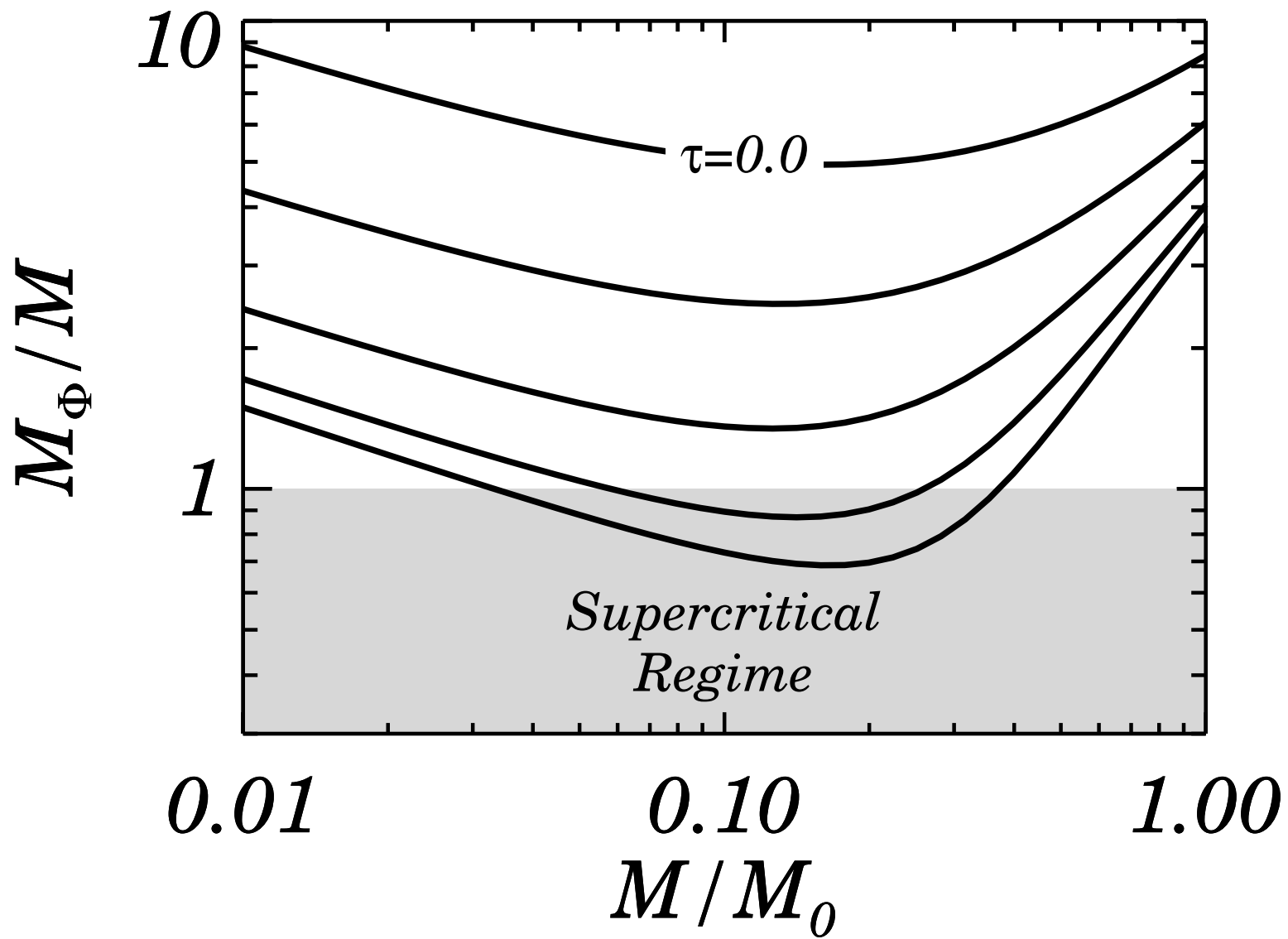
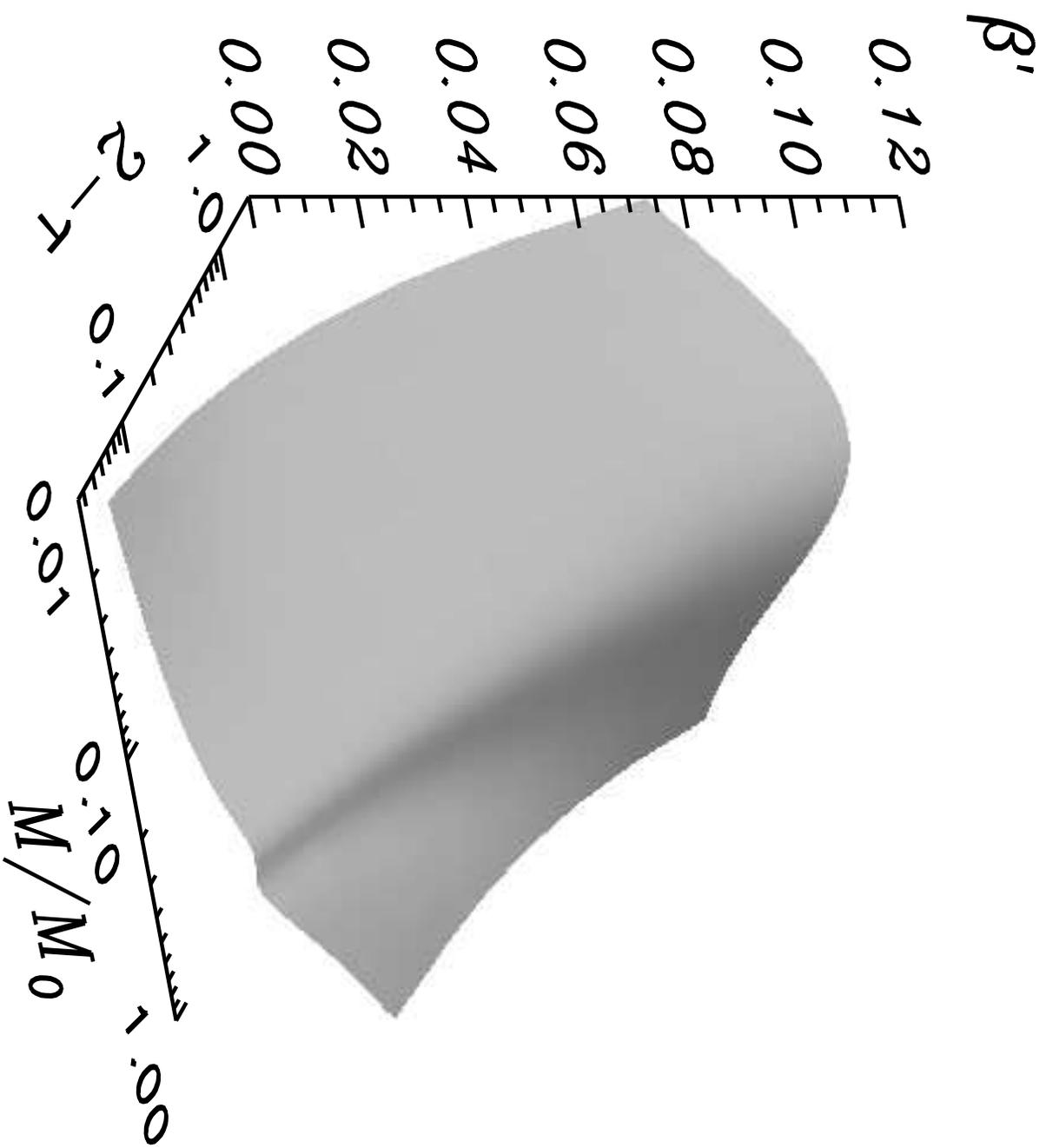
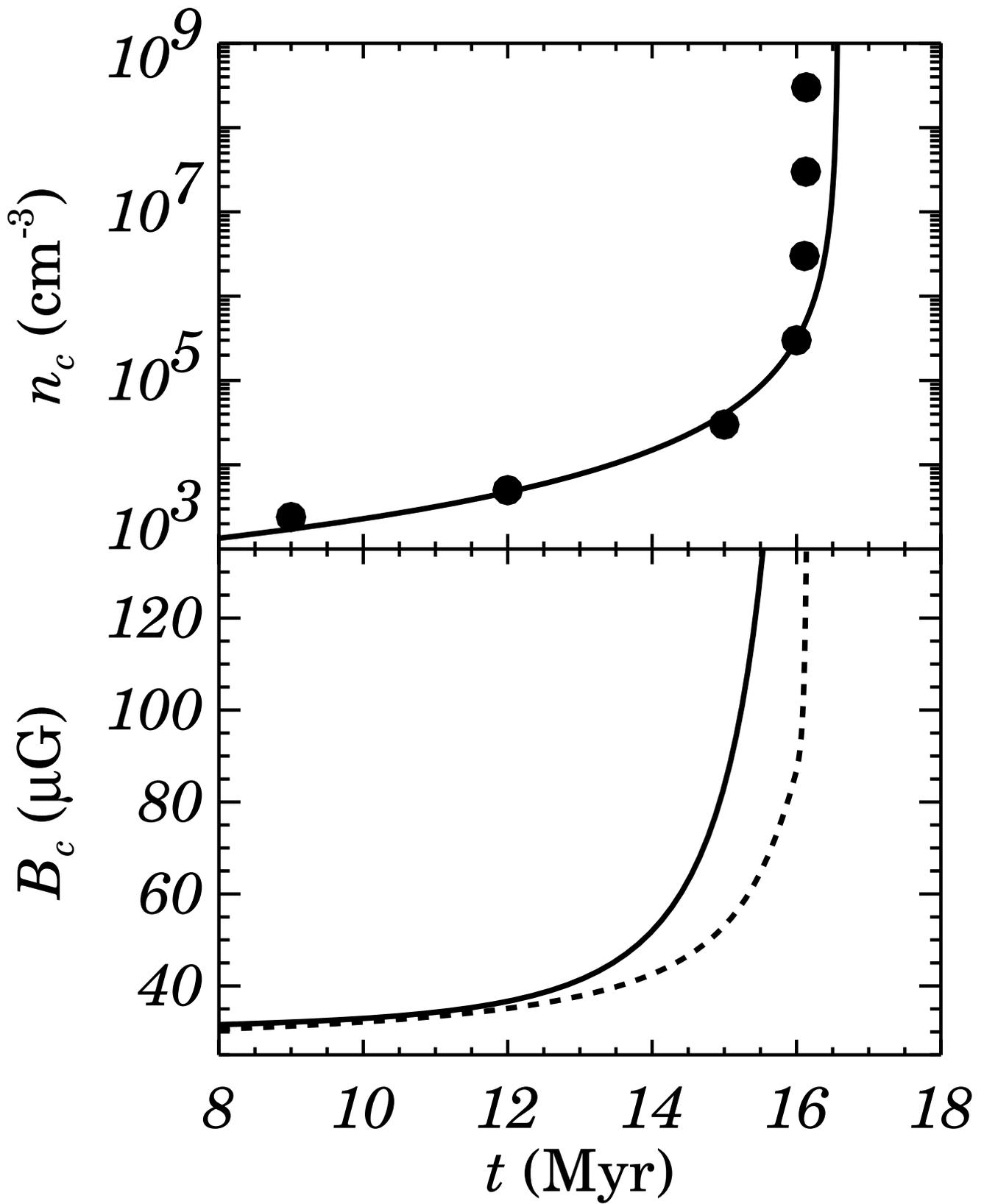
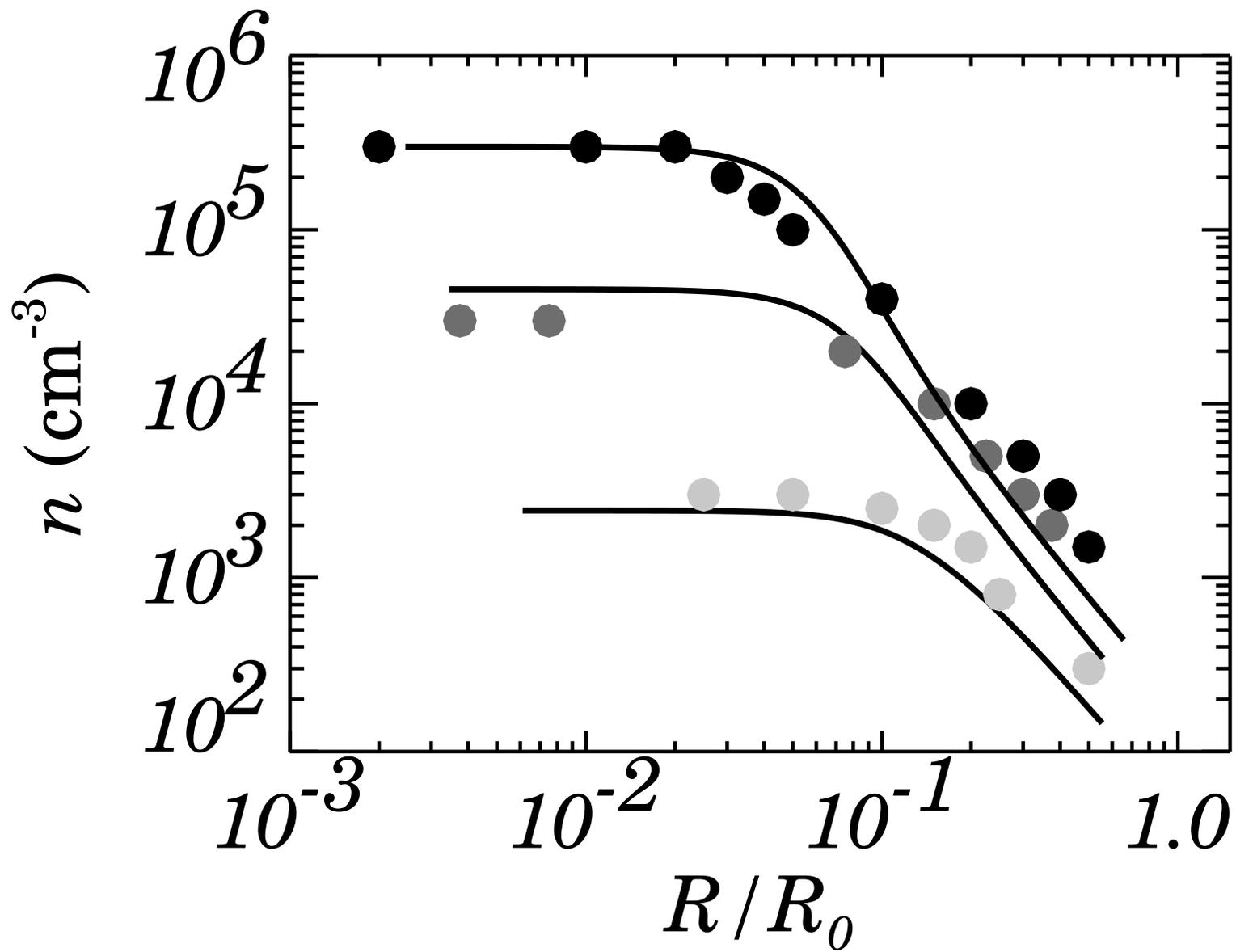
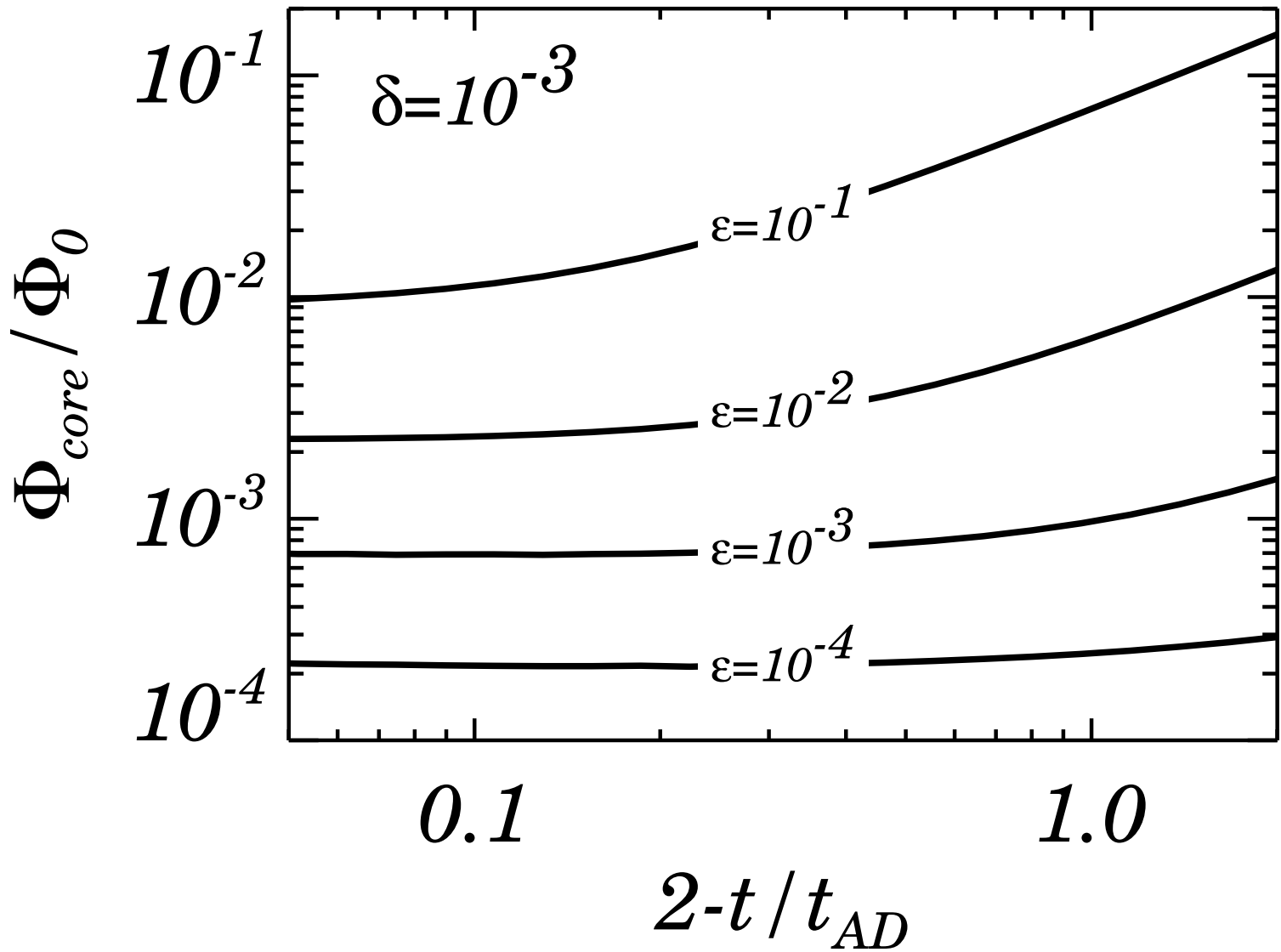
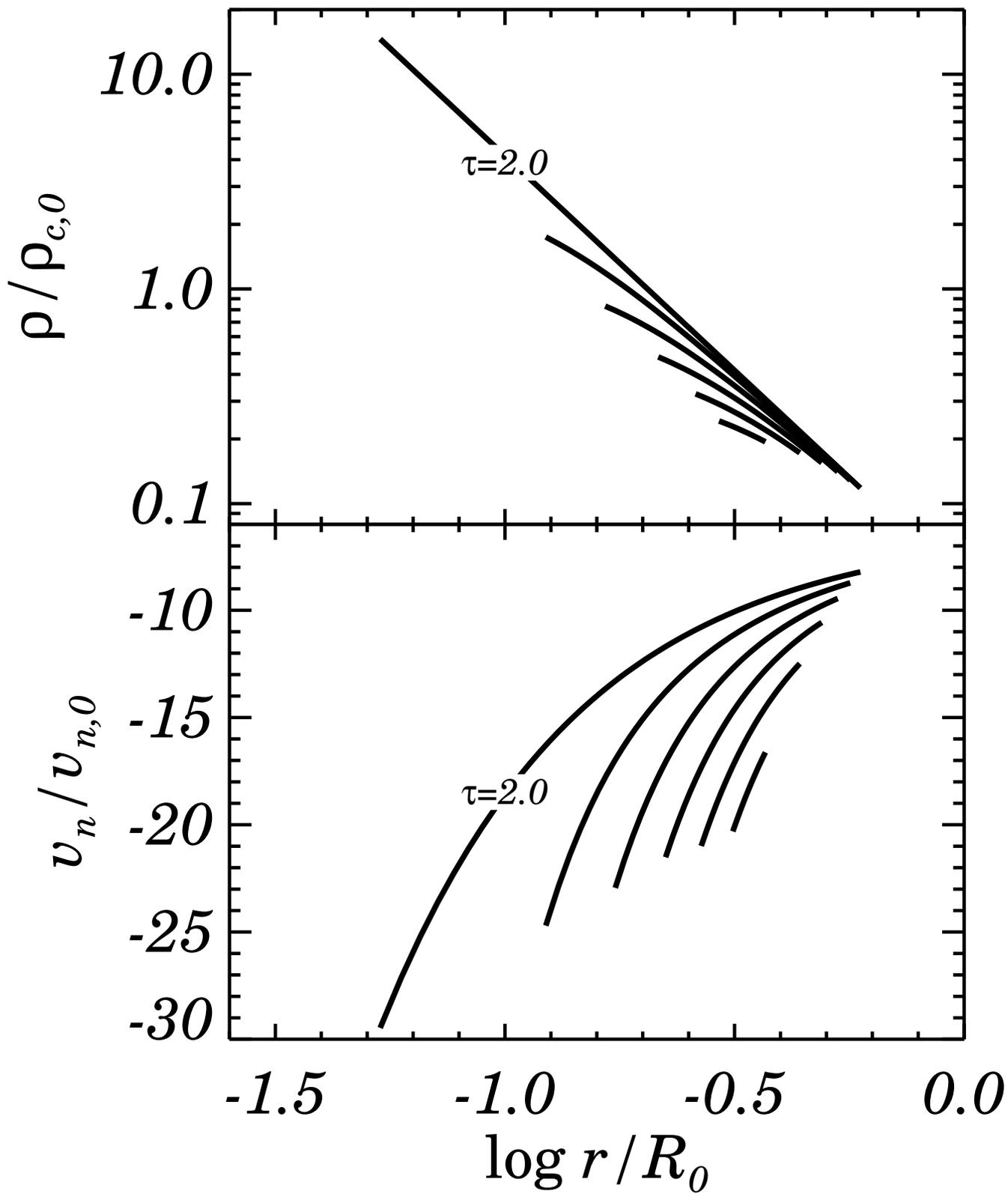
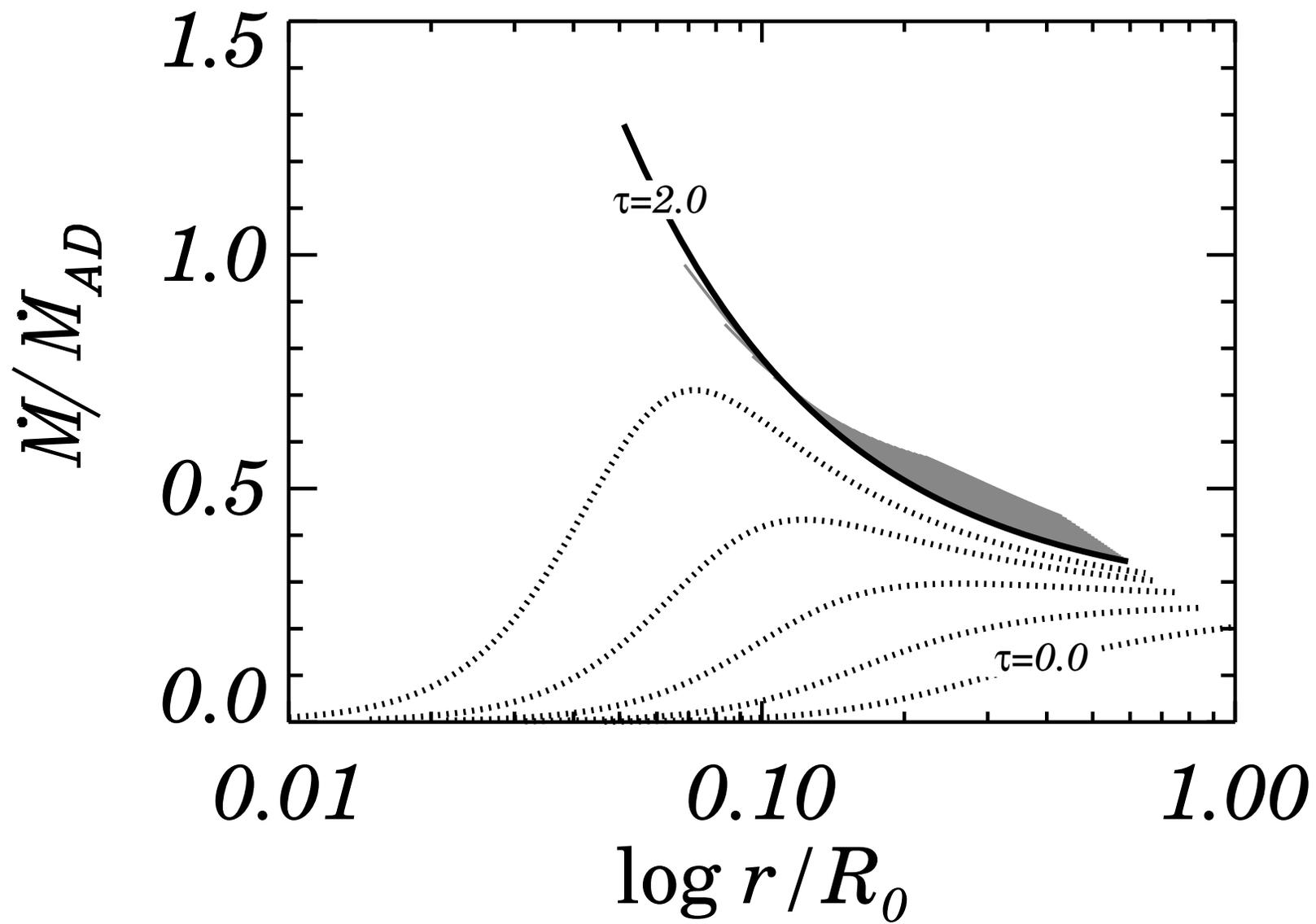
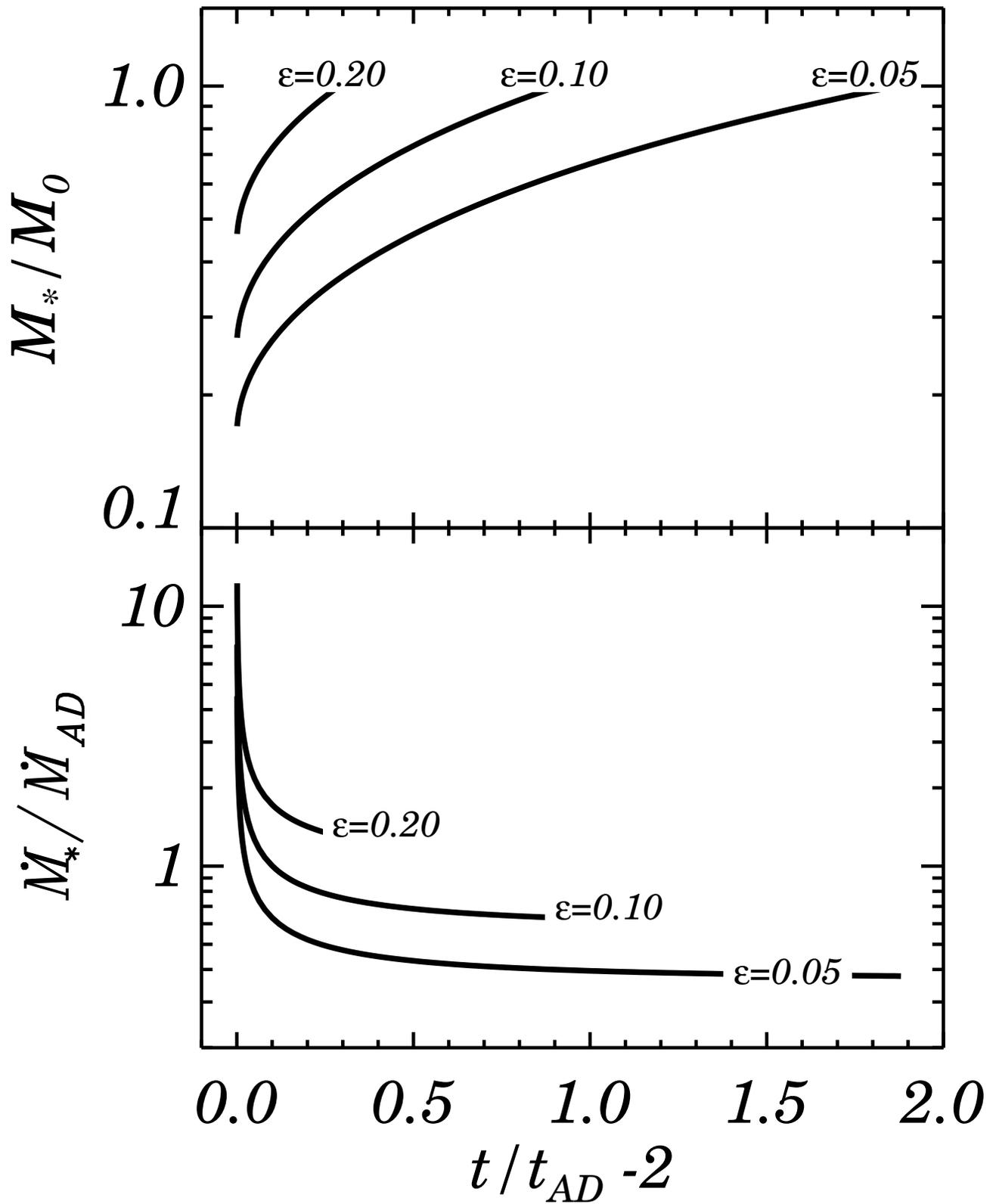
\end{document}